\documentclass[fleqn,12pt]{wlscirep}
\usepackage[utf8]{inputenc}
\usepackage[T1]{fontenc}
\usepackage{newtxtext,newtxmath}
\usepackage{float}
\usepackage{placeins}

\usepackage{gensymb}
\usepackage{soul}
\usepackage{enumitem}
\usepackage{comment}
\usepackage[numbers,sort&compress]{natbib}
\usepackage{setspace}
\usepackage{lineno}

\newcommand{\egcite}[1]{[e.g.,~\citealp{#1}]}

\title{Predicting Beyond Training Data via Extrapolation versus Translocation: AI Weather Models and Dubai's Unprecedented 2024 Rainfall}

\author[1,*]{Y. Qiang Sun}
\author[1,2,*]{Pedram Hassanzadeh}
\author[1]{Tiffany Shaw}
\author[3]{Hamid A. Pahlavan}
\affil[1]{University of Chicago, Department of the Geophysical Sciences, Chicago, IL 60637, USA}
\affil[2]{University of Chicago, Committee on Computational and Applied Mathematics, Chicago, IL 60637, USA}
\affil[3]{NorthWest Research Associates, Boulder, CO 80301, USA}

\affil[*]{qiangsun@uchicago.edu and pedramh@uchicago.edu}

\begin{abstract}

Artificial intelligence (AI) models have transformed weather forecasting, but their skill for gray swan extreme events is unclear. Here, we analyze GraphCast and FuXi forecasts of the unprecedented 2024 Dubai storm, which had twice the training set’s highest rainfall in that region. Remarkably, GraphCast accurately forecasts this event 8 days ahead. FuXi forecasts the event, but underestimates the rainfall, especially at long lead times. GraphCast’s success stems from “translocation”: learning from comparable/stronger dynamically similar events in other regions during training via a global effective receptive field. Evidence of “extrapolation” (learning from training set's weaker events) is not found. Even events within the global distribution’s tail are poorly forecasted, which is not just due to data imbalance (generalization error) but also spectral bias (optimization error). These findings demonstrate the potential of AI models to forecast regional gray swans and opportunity to improve them through understanding the mechanisms behind their successes and limitations.

\vspace{20mm}

\end{abstract}

\newpage
\begin{document}

\flushbottom
\maketitle

\thispagestyle{empty}

\newpage

\begin{doublespace}

\section{Introduction}

Artificial intelligence (AI) weather models such as FourCastNet~\cite{pathak2022fourcastnet}, Pangu~\cite{bi2023accurate}, GraphCast~\cite{lam2022graphcast}, FuXi~\cite{chen2023FuXi}, AIFS~\cite{lang2024aifs}, NeuralGCM \cite{kochkov2024neural}, and GenCast~\cite{price2025probabilistic} have demonstrated remarkable skill in forecasting global weather patterns on short- to medium-range timescales. These AI models consistently match or even surpass the performance of state-of-the-art physics-based numerical weather prediction (NWP) models. The key advantages of these AI weather models—which include higher accuracy, orders-of-magnitude faster forecasting speed, and significantly lower computational cost~\egcite{pathak2022fourcastnet,bi2023accurate,chen2023fengwu,mahesh2024huge2,bracco2024machine,ben2024rise}—have captured the attention of the research community and are beginning to gain traction within operational weather centers~\cite{mcgovern2017using,sun2020applications,schultz2021can,kashinath2021physics,clark2024advancing,lang2024aifs}.


Despite the overall success of these AI models in short- and medium-range weather forecasting based on common metrics, there are concerns about their ability to reliably forecast the rarest yet most extreme weather events, particularly the so-called ``gray swans''~\cite{watson2022machine, lam2023learning,rasp2024weatherbench,mcgovern2025extreme}. First used in climate science by Lin and Emanuel~\cite{lin2016grey}, a gray swan refers to a rare extreme weather event that is physically possible, but due to its long return period, has not been seen in the available historical record. Recently, Sun~et al.~\cite{sun2024can} used this term in the context of AI weather/climate modeling for physically possible extreme events that are so rare they are absent from the \textit{training dataset}. 
Forecasting such events accurately requires AI models to generalize beyond their training data distribution, i.e., to extrapolate, which potentially necessitates learning underlying physical processes rather than merely recognizing patterns. The AI models' ability to forecast out-of-distribution events, such as gray swans, poses a significant test for these models. This question also has major social implications given the large impact of these strongest weather events on the one hand and the rapid rise of AI-driven operational weather forecasting on the other \cite{dueben2018challenges,chantry2021opportunities,watson2022machine,camps2025artificial}.

Several case studies have begun to explore the ability of AI models to forecast individual extreme events~\cite{charlton2024ai,pasche2025validating,duan2024ai,xu2024evaluating,bano2025ai}. However, further analyses are needed to quantify to what extent these test cases were out-of-distribution regionally and globally. Moreover, much of the existing literature focuses on evaluating the models' forecast skills without delving into the mechanisms that underpin their success or failure. Understanding differences in the performance of models in forecasting certain extreme events is critical for further advancing AI weather forecasting systems. 

For example, the error of deep learning models has three sources: approximation, generalization, and optimization \cite{shalev2014understanding,lu2021learning}. The difficulty in learning extreme weather events is often attributed to their rarity, which leads to statistical learning's classic ``data imbalance'' problem, a form of \textit{generalization error} \cite{he2013imbalanced,krawczyk2016learning}. Gray swans face the ultimate (asymptotic limit of) data imbalance problem. However, the large amplitudes of these events also make them prone to what is known as the blurring (or double penalty) problem in the AI for climate literature \cite{rasp2024weatherbench,subich2025fixing}, which is, in fact, the classic ``spectral bias'' in deep learning \cite{rahaman2019spectral}. Theoretical work has shown that spectral bias, i.e., the challenges of neural networks in learning small scales (high wavenumber), is an \textit{optimization error} \cite{cao2019towards,xu2024overview}. Given that errors from data imbalance and spectral bias have different roots and thus different potential remedies, it is critical to understand their roles in investigations of AI weather models' performance for extreme events, and, in particular, for gray swans.


With these questions in mind, here, we examine the medium-range forecast skills of two state-of-the-art AI weather models, GraphCast and Fuxi, for the April 2024 extreme rainfall in Dubai and a few other extreme rainfall episodes in the Northern Hemisphere. We chose the Dubai event for two reasons. First, as shown later, this record-breaking event had nearly twice the highest rainfall in this region in the training set of these models. In addition to the effects of rarity and spectral bias on the AI-driven forecasts of this gray swan event, we investigate these models' ability to learn from the extreme events in other regions by calculating the effective \textit{receptive field} of their neural network. These analyses provide insight into the strengths and weaknesses of these models in forecasting out-of-distribution extreme events, thus informing future developments in AI-based weather prediction systems. 

Second, while recent advances have demonstrated the skill of AI models in nowcasting ~\cite{ravuri2021skilful,espeholt2022deep,zhang2023skilful} or short-term forecasting~\cite {rasp2024weatherbench,price2025probabilistic,lang2024aifs} of rainfall, their performance for extreme rainfall events, especially out-of-distribution events, remains unexplored. 
This gap is critical, as such events are known to be difficult to predict with traditional NWP models and often have severe societal impacts. Dubai's April 2024 event serves as an example: as an arid region not prepared for such extremes, the city experienced widespread flooding and infrastructure damage, with estimated insurance losses of approximately US\$3 billion and five reported fatalities~\cite{jba2024dubai}. This highlights the urgent need for improved forecasting capabilities and the rapid public dissemination of reliable early warnings. The possibility of AI models to skillfully predict extreme rainfall several days in advance, as examined in this study, combined with their computational efficiency and public availability, offers a promising new pathway for delivering early warnings through multi-model, large-ensemble forecasts~\cite{camps2025artificial}.

\section{Skill of AI models for the unprecedented rainfall event} \label{sec:skill}

\begin{figure}[ht]
\centering
\includegraphics[width=1.0\linewidth]{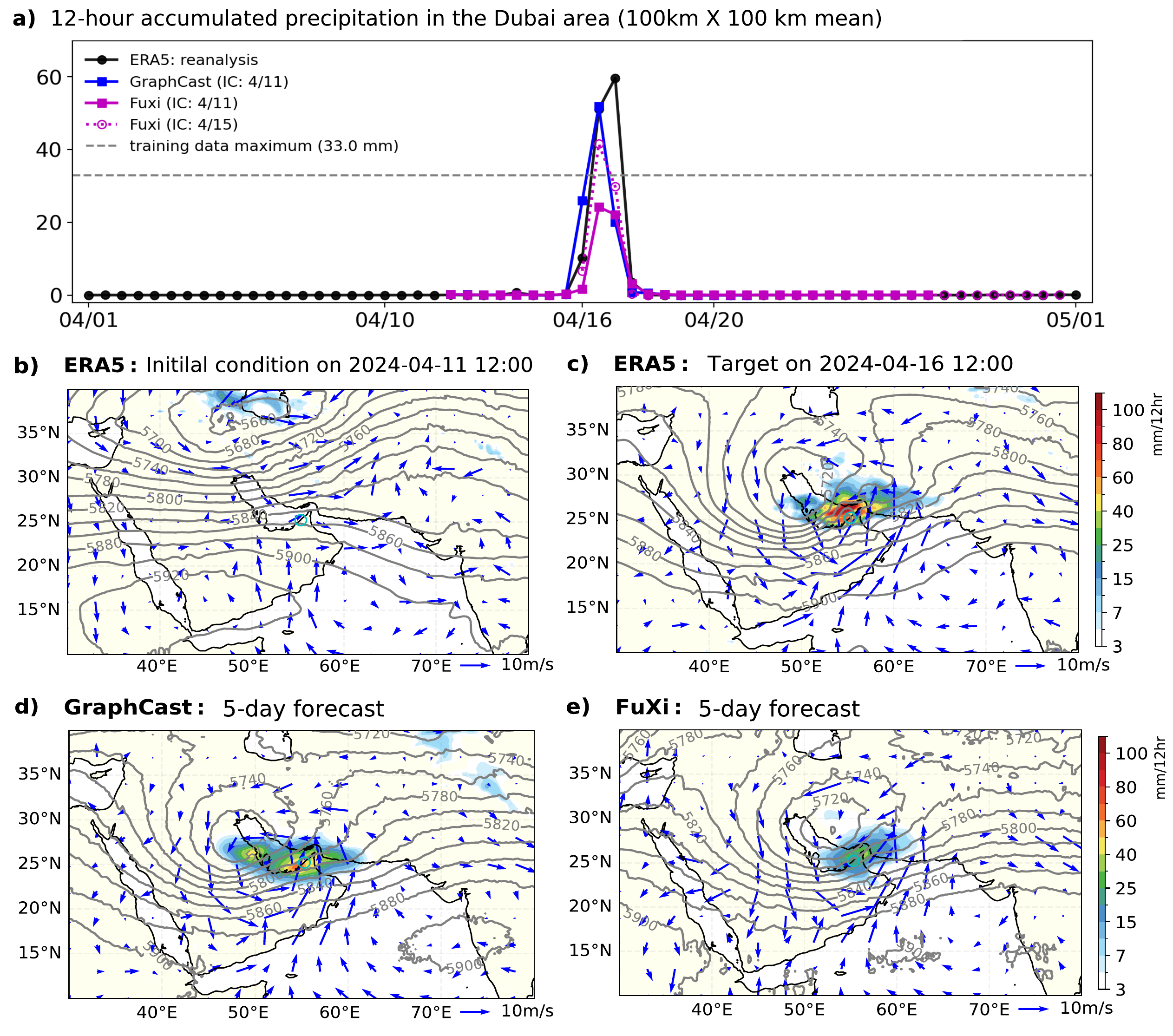}
\caption{\textbf{Forecasting Dubai's 2024 extreme rainfall event using AI weather models.} \textbf{(a)} Time series of 12-hour accumulated precipitation averaged around Dubai, comparing ERA5 reanalysis (black line) with forecasts from GraphCast (blue line, 5 days ahead) and FuXi (magenta line, 5 days ahead; dashed line, 1 day ahead). The horizontal dashed line indicates the maximum 12-hour accumulated precipitation (33~mm) in this region ever observed in the period of 1979-2021, covering the training period for both GraphCast and Fuxi. \textbf{(b)} Circulation pattern and rainfall over the Arabian Peninsula on April 11th, 2024, at 12:00 UTC (5 days before the event), showing 500 hPa geopotential height (contours),  850 hPa wind vectors (arrows), and 12-hour accumulated precipitation (shading), all from  ERA5 reanalysis. \textbf{(c)} Same as (b), but for 5 days later, April 16th, 2024, at 12:00 UTC, when the instantaneous rainfall value is close to its peak. \textbf{(d)} Same as (c), but for a 5-day forecast by GraphCast initialized on April 11th (panel b). \textbf{(e)} Same as (d), but for a 5-day forecast by FuXi. See Figure~S2 for circulation forecasts with Pangu and FourCastNetV2 AI weather models, which do not provide precipitation.}
\label{fig: Dubai-prediction}
\end{figure}

\begin{figure}[ht]
\centering
\includegraphics[width=1.0\linewidth]{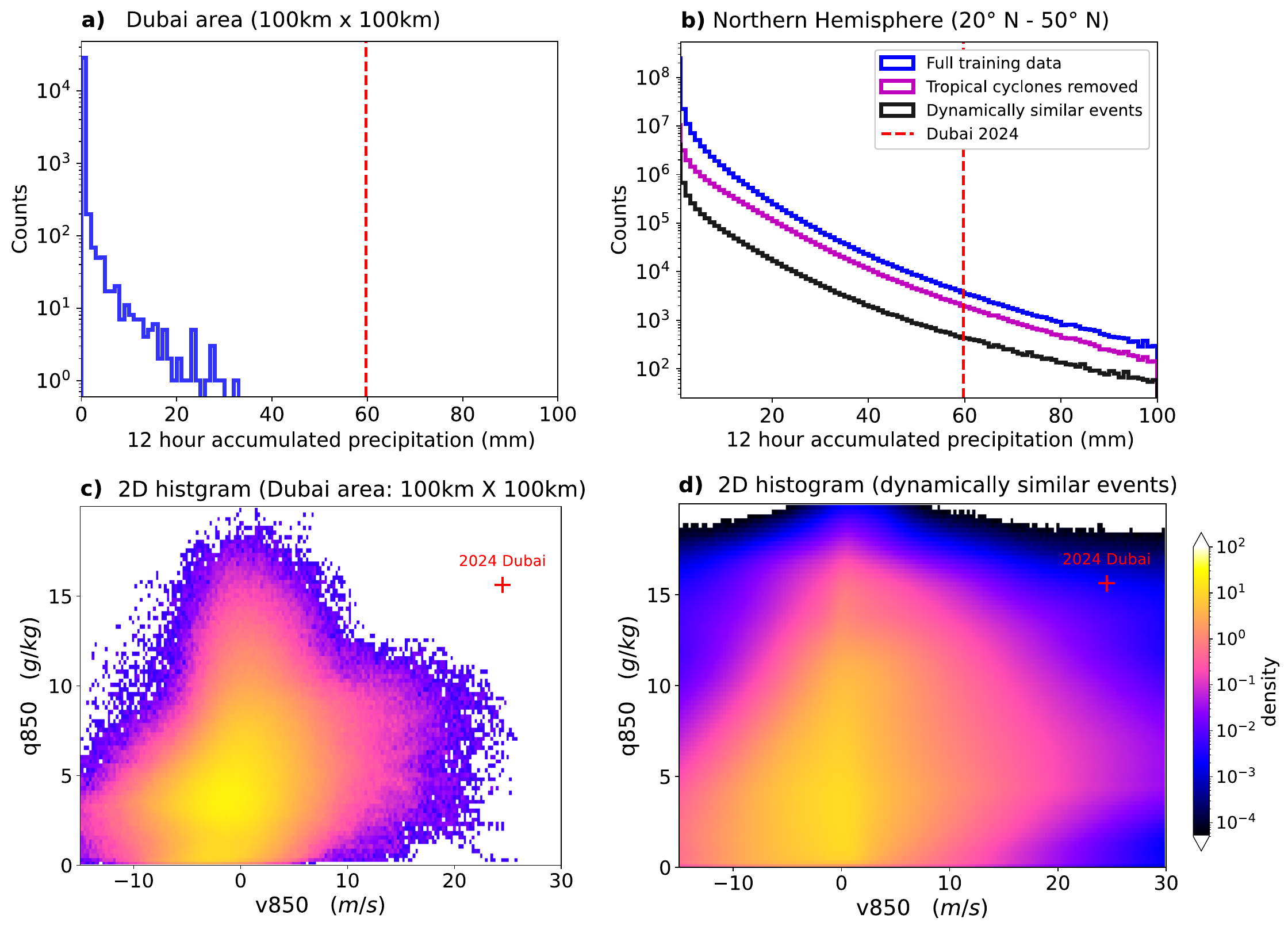}
\caption{\textbf{The Dubai rainfall event as a regional gray swan}. \textbf{(a)} Histograms of 12-hour accumulated precipitation within a $100$~km $\times 100$~km box around Dubai in the training dataset for GraphCast and FuXi models (ERA5, 1979-2017). \textbf{(b)} Same as (a), but focused on the 20°N-50°N latitude band, considering the full training set and the same dataset but with tropical cyclones excluded or only dynamically similar events (extratropical cyclones) included; see Methods for details. The vertical lines indicate the 12-hour accumulated precipitation during the 2024 Dubai extreme event. \textbf{(c)} Moisture transport around Dubai in the training set is shown as a 2D histogram of 850 hPa meridional wind   (v850) and specific humidity at (q850). The red '+' symbol marks the values of the 2024 Dubai event (April 15-16). \textbf{(d)} Same as (c), but for all dynamically similar events in the 20°N-50°N latitude band. }

\label{fig: Dubai-PDF}
\end{figure}

Dubai's extreme rainfall event that started around April 15, 2024, is evident in the observation-derived ERA5 reanalysis data, with a peak 12-hour accumulated value of 60~mm on April 16 even after averaging over a $100~\text{km} \times 100~\text{km}$ box centered around the city (Figure~\ref{fig: Dubai-prediction}a). This value is almost twice the highest precipitation in the training set of the AI models (33~mm), making this a gray swan event (see the histogram in Figure~\ref{fig: Dubai-PDF}a). Yet, with a 5-day lead time, both GraphCast and FuXi, which provide precipitation forecasts, accurately capture the timing of the event, with an error margin of one or two 6-hourly time steps. Notably, GraphCast's 5-day forecast predicts a maximum precipitation value of 55~mm (on April 16), which is close to that of ERA5 (Figure~\ref{fig: Dubai-prediction}a). Although GraphCast's forecast of the total accumulated rainfall during this event (15-18 April) is still lower than what is seen in ERA5 (by $\sim 15 \%$), the model's ability to forecast the timing, location (Figure~\ref{fig: Dubai-prediction}c-d), and the overall amplitude of such an unprecedented event is somewhat surprising. This finding might appear to suggest GraphCast's ability to generalize out-of-distribution. Note that even the 8-day forecast skill of GraphCast is reasonable, capturing the circulation and rainfall patterns (Figure~S1) and producing a peak rainfall of $\sim 41$~mm around Dubai on April 16.  

While FuXi also captures the event's timing, location, and circulation pattern (Figure~\ref{fig: Dubai-prediction}e), its 5-day forecast significantly underestimates the peak, producing values below the training maximum of 33~mm. This might appear to suggest Fuxi's inability to generalize out-of-distribution. Before further analyzing the difference between the performance of these two models for this extreme event, below we briefly discuss two issues: the fidelity of precipitation in ERA5 and the meteorological conditions through the evolution of this event. The latter discussion will help with understanding some of the later analyses in the paper.

\FloatBarrier

\FloatBarrier


The representation of precipitation in ERA5 is known to have shortcomings, mainly due to its coarse resolution and the lack of direct assimilation of relevant observations~\cite{tang2020have,lavers2022evaluation}. However, for the purpose of this study, it suffices (and is in fact necessary) to examine the AI weather models in the ``ERA5 world'': since they are trained on ERA5 data (including precipitation), we test them against ERA5 (including its precipitation) as the ground truth. Note that the shortcomings of ERA5 precipitation can be addressed in practice by training or fine-tuning the AI models on observation-derived datasets, such as IMERG~\cite{yuval2024neural}. 

\begin{figure}[ht]
\centering
\includegraphics[width=1.0\linewidth]{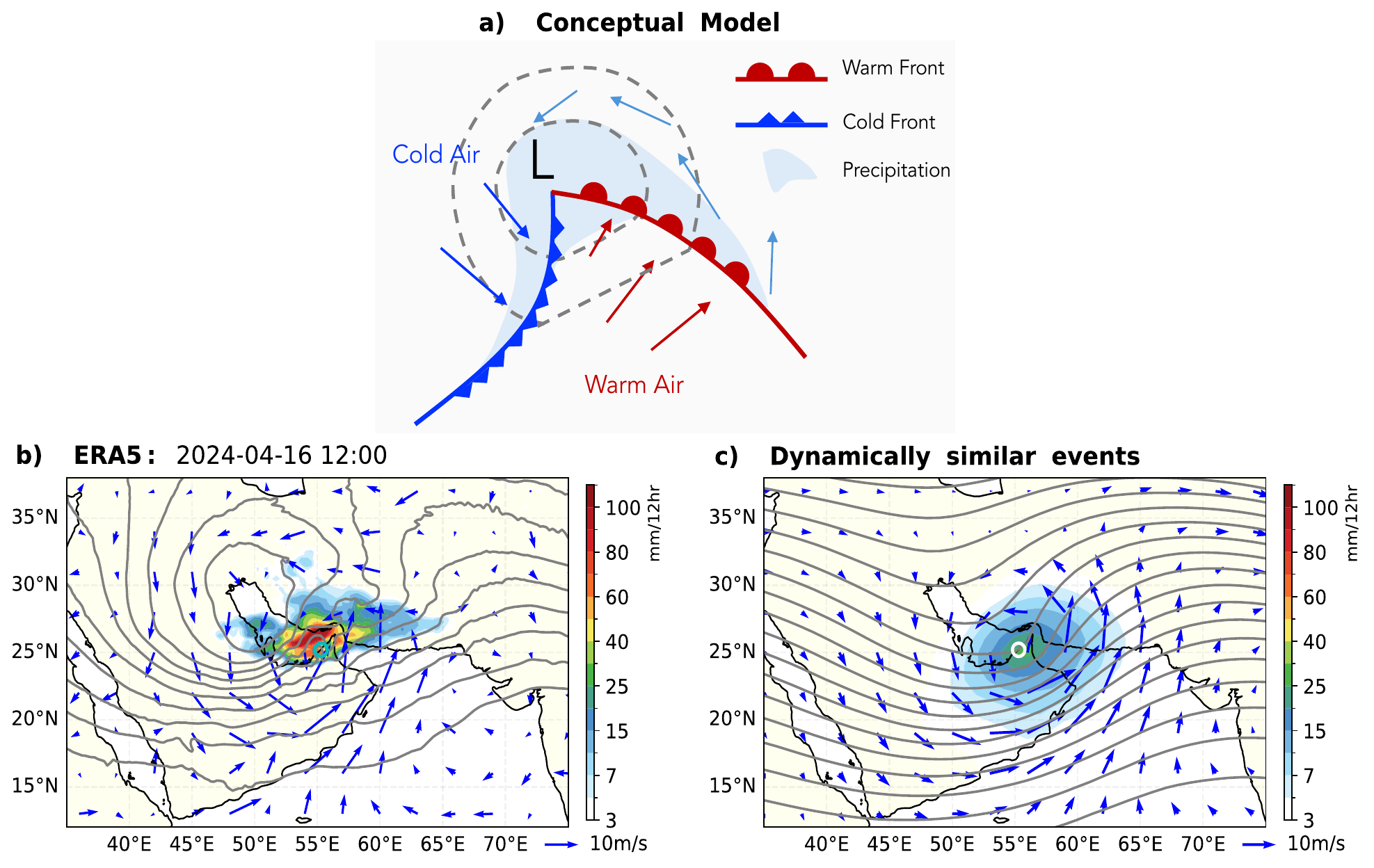}
\caption{\textbf{Dynamically similar events in other regions within the training dataset.} \textbf{(a)}  Conceptual model of an extratropical cyclone-driven heavy precipitation event, similar to the 2024 Dubai event. \textbf{(b)} Circulation and precipitation of the 2024 Dubai event in ERA5 (same as Figure~\ref{fig: Dubai-prediction}c). \textbf{(c)} Composite of dynamically similar events that have precipitation comparable to or even stronger than the Dubai event, identified from the training set in ERA5 over 20\degree $-$50\degree \,latitude band of the Northern Hemisphere using a Lagrangian tracking algorithm (see Methods for more details). The patterns are centered on the city of Dubai for the composite.}
\label{fig: composite}
\end{figure}

The textbook conceptual model of a extratropical cyclone is characterized by the convergence of warm and cold air masses and the associated precipitation near frontal boundaries (Figure~\ref{fig: composite}a). All elements of this conceptual model are present in the Dubai event (Figure~\ref{fig: composite}b). Briefly, the meteorological conditions five days before the event include a trough extending over the northern Arabian Peninsula, accompanied by southerly winds at 850 hPa, and light rainfall over Iran (Figure~\ref{fig: Dubai-prediction}b). During the peak of the event on April 16, a fully developed cut-off low-pressure system is evident in the 500 hPa height field (z500), centered over the Persian Gulf (Figure~\ref{fig: Dubai-prediction}c). The intensified meridional winds at 850 hPa (v850) facilitated substantial moisture advection into the region, resulting in concentrated areas of heavy rainfall, as indicated by 12-hour accumulated precipitation exceeding 100 mm in localized spots near Dubai. This value exceeded the climatological accumulated annual rainfall for Dubai, highlighting the severity of the event. Further analyses of the histograms of several key variables show that while the values of low-level meridional wind and relative humidity (v850 and q850) were at the tail of the distribution (of the training set) around Dubai, they were not unprecedented (Figure~S3). What was unprecedented and substantially out-of-distribution for this region was the northward moisture transport, which can be seen in the joint histogram of v850 and q850 (Figure~\ref{fig: Dubai-PDF}c). Note that z500 (or its anomaly) and mean sea-level pressure (mslp) were well in-distribution for this event (Figure~S3).   

\FloatBarrier

It should be noted that The European Center for Medium-Range Weather Forecasts (ECMWF) found their NWP model to also forecast this event fairly accurately\footnote{https://www.ecmwf.int/en/newsletter/180/news/unprecedented-rainfall-united-arab-emirates}. As the focus of this paper is on the ability of AI weather models in forecasting out-of-distribution precipitation events, we have not analyzed or presented the NWP results.


\section{Factors contributing to the success of AI forecasts for this unprecedented  event} \label{sec:success}
For an AI model to successfully forecast a regional out-of-distribution event such as Dubai's rainfall there are two possible mechanisms:

\begin{enumerate}[itemsep=0pt, parsep=0pt]
\item Extrapolation – The model learns and generalizes out-of-distribution from weaker (thus more frequent), local, similar events that it has seen in the training set.
\item Translocation – The model learns from comparably strong or stronger dynamically similar events occurring in other regions in the training set.
\end{enumerate}
Extrapolation is the commonly hypothesized mechanism for the potential success of AI weather models with gray swans. Translocation is a mechanism that emerged from a recent study by some of the current authors and their collaborators \cite{sun2024can}. That study, which was inspired by this analysis of the Dubai unprecedented rainfall event, trained versions of another AI weather model (FourCastNet) with modified training sets. It was shown that the AI model cannot forecast Category 5 tropical cyclones (TCs) if it has not seen any Category 3-5 TCs in its training set, clearly demonstrating a lack of extrapolation. However, it was also shown that the version trained without Category 3-5 TCs in only one ocean basin (Atlantic or Pacific) has skill forecasting Category 5 TCs in that basin, suggesting that the AI model can generalize across tropical basins (i.e., translocate). They further argued that translocation only works among events with similar dynamics, which is manifested to the AI model through the relationship between variables in the entire state vector.  

Below, we will first present an in-depth analysis of learning from translocation followed by investigating extrapolation. Note that the two mechanisms are not mutually exclusive. 

\subsection{Learning via translocation of stronger dynamically similar events}




As shown in Figure~\ref{fig: Dubai-PDF}a and c, this event was a \textit{local} gray swan in terms of the training dataset's distribution of 12-hour accumulated precipitation around Dubai, resulting from an unprecedented northward moisture transport. However, across the broader Northern Hemisphere, precipitation events with a similar or higher magnitudes exist in the training set (Figure~\ref{fig: Dubai-PDF}b). Recognizing that some of the highest precipitation values are associated with TCs moving into the extratropics, we conduct additional analyses to exclude TC-induced precipitation (pink line in Figure~\ref{fig: Dubai-PDF}b). Furthermore, we isolate precipitation only from events that have a meteorological driver similar to those of the Dubai event, that is, a extratropical cyclone. Details of this analysis can be found in the Methods section. Briefly, using a Lagrangian tracking algorithm, we have identified dynamically similar events with precipitation that is comparable or higher than the Dubai event in the training set. The circulation and precipitation composites of these events closely resemble those of the Dubai event (Figure~\ref{fig: composite}c), albeit with reduced amplitude due to the smoothing effect of averaging many cases.

Figure~\ref{fig: Dubai-PDF}b shows that many events with dynamics similar to that of the Dubai event but with comparable or higher precipitation existed in other regions of the training set of the AI models. The same is true for moisture transport, as demonstrated in the 2D histogram of v850 and q850 (Figure~\ref{fig: Dubai-PDF}d). This shows that the key dynamical process of this event was ``in-distribution'' when the entire Northern Hemisphere is considered. 

These results suggest that while the 2024 event was unprecedented for the Dubai area, many dynamically similar events existed elsewhere in the Northern Hemisphere in the training set, providing potential learning opportunities for such a ``regional'' gray swan via translocation. Building on the findings of Sun et al.~\cite{sun2024can}, we speculate that this is at least one way for GraphCast to forecast the Dubai event so well. Note that demonstrating translocation unambiguously following the framework of \cite{sun2024can} requires training the AI model from random weights with a modified data set (e.g., with all samples containing stronger dynamically similar events removed from other regions). However, this procedure is computationally very expensive for GraphCast and is not pursued here. Instead, in this work, we leverage a computational tool based on deep learning theory to understand if GraphCast could use spatially non-local information as it learned about Dubai area's precipitation.    


For a convolutional neural network, the ``receptive field'' is the region of the input that the network uses to predict the output~\cite{luo2016understanding,pahlavan2024importance}. Theoretically, the receptive field can be calculated knowing the depth and convolution kernel size of the network and provides information on whether spatially non-local information of the input is used by the network. For any architecture, this can be estimated using the effective receptive field (ERF); see Pahlavan et al.~\cite{pahlavan2024importance} for an example in climate modeling. More specifically, ERF is computed by backpropagating a signal from the output layer to the input layer. ERF quantifies the influence of different input regions on a given output (specific variable at a specific location)—similar to what adjoint methods do, which for example, have been recently used to estimate forecast sensitivity in AI models to initial conditions~\cite{vonich2024predictability, bano2025ai}. See Methods for details of the ERF computation. 

\begin{figure}[ht]
\centering
\includegraphics[width=1.0\linewidth]{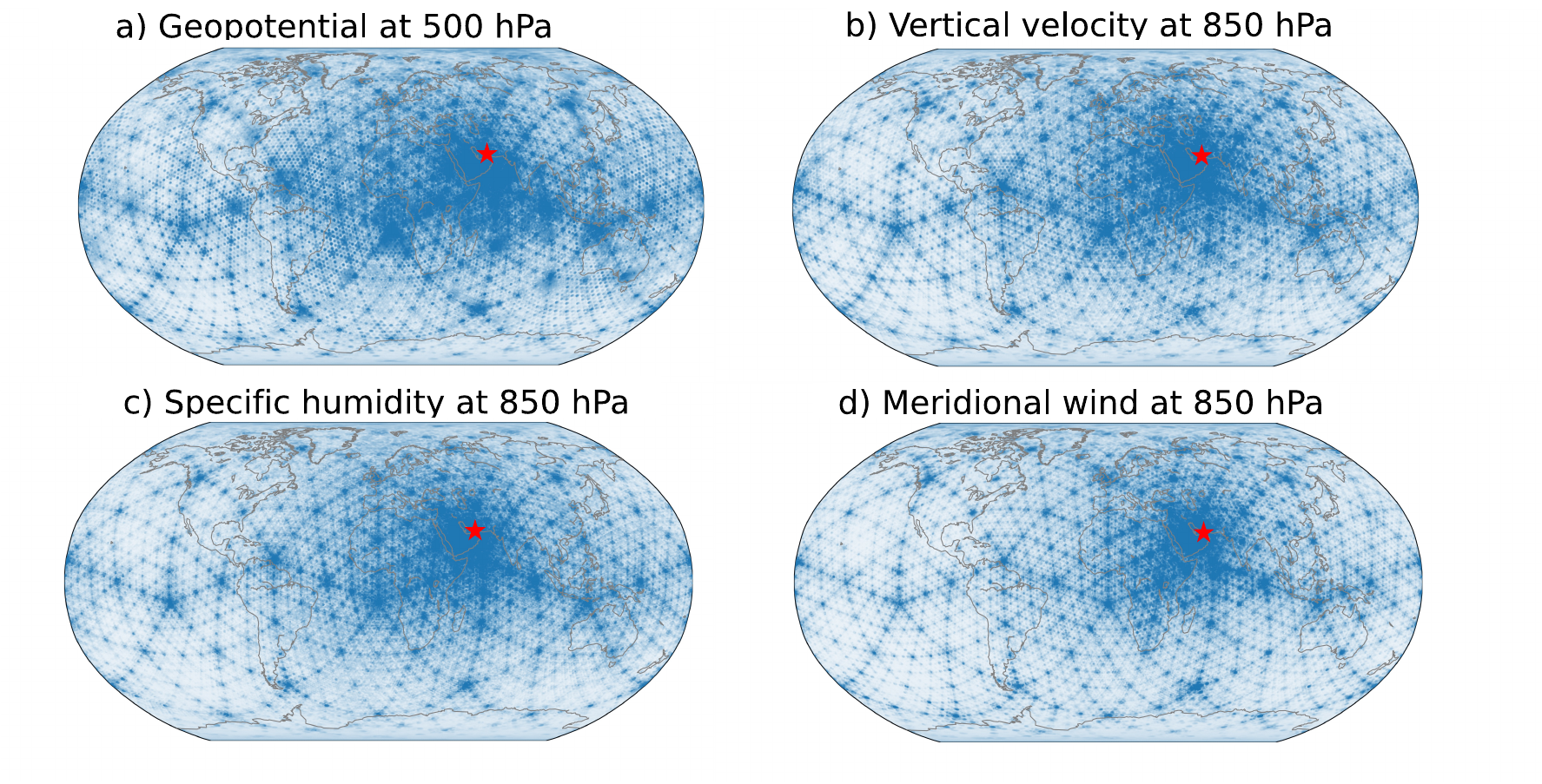}
\caption{\textbf{Effective receptive field (ERF) of GraphCast for forecasting precipitation (6 hours later) in the Dubai area.} Four selected variables are shown: \textbf{(a)} geopotential at 500~hPa, \textbf{(b)} vertical velocity at 850~hPa, \textbf{(c)} specific humidity at 500~hPa, and \textbf{(d)} meridional wind at 850~hPa. The forecast is initialized at 12:00 UTC on April 15, 2024, with ERF calculated after one time-step (6 hours). The star symbol in each panel shows the location of Dubai. GraphCast uses two previous time steps for predictions; we show results for the later time step, as the earlier step exhibits a similar but weaker pattern (not shown). The maps are scatter plots where the point size represents ERF's amplitude, with darker blue indicating stronger influence on GraphCast’s precipitation predictions over Dubai. The model’s multi-icosahedral mesh structure is distinctly visible. See Methods for details of ERF calculations.}
\label{fig: ERF}
\end{figure}

The ERF of GraphCast for one time-step (6-hourly) forecast of precipitation around Dubai is shown in Figure~\ref{fig: ERF} for selected variables that play a role in the dynamics of extratropical cyclones-driven precipitation. 
The analysis shows that while GraphCast mainly relies on local information, it also exhibits a ``global'' ERF that aligns with the multi-icosahedral mesh of GraphCast. The ERF shows substantial non-locality that, in almost all variables analyzed (e.g.,  850~hPa wind and specific humidity, and z500), covers a large area from the east Atlantic to the east Pacific. We note that this is the almost instantaneous ERF (from a one-time step, 6-hourly forecast); it is {\it not} a result of the advection and propagation of waves and teleconnection to the Dubai area over time. The highly non-local ERF suggests that during training, which is done using two 6-hourly time steps, GraphCast could have seen other (potentially stronger) dynamically similar events in other regions when learning to forecast precipitation around Dubai. This lends further support to learning via translocation. More research is needed to unambiguously confirm this hypothesis and gain deeper insight into the non-local learning process of AI weather model and how it can be leveraged with other physical constraints considered (see Section~\ref{sec:summary} for more discussions). Examining the ERF of FuXi and other models is left for future work.       


\FloatBarrier

\subsection{Learning via extrapolation from weaker local events}

The earlier analysis suggest that GraphCast might have learned to forecast the Dubai event, a regional gray swan, via translocation. However, this does not exclude the possibility of learning from extrapolation from weaker events. 

\begin{figure}[ht]
\centering
\includegraphics[width=\linewidth]{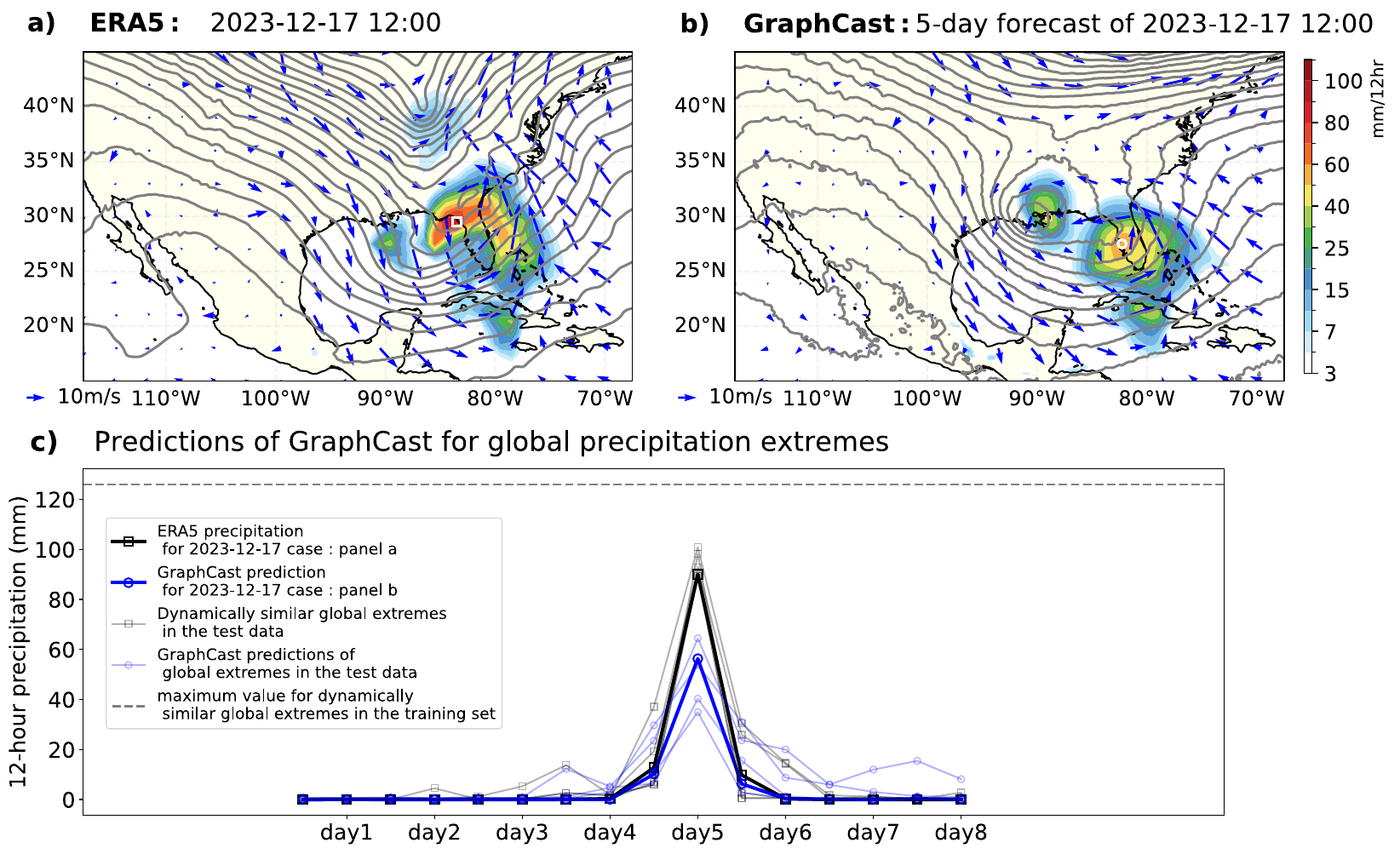}
\caption{\textbf{GraphCast's limitations in accurately predicting highest amplitude precipitation extremes of the Northern Hemisphere.} \textbf{(a)} similar to Figure~\ref{fig: Dubai-prediction}a, but for one of the top 5 extratropical cyclone-driven extreme precipitation events of the test set (2020-2023). This event happened in December 2023 over Florida, USA, with 12-hour accumulated precipitation reaching 90~mm.  \textbf{(b)} Same as (a), but for a 5-day forecast with GraphCast. \textbf{(c)} Time series of 12-hour accumulated precipitation for the top 5 dynamically similar events (see Figure~S4 for the circulation and precipitation patterns of the other 4 cases). These events all have 12-hour accumulated precipitation exceeding 90~mm. The black and grey lines represent the ``truth'' in the ERA5 reanalysis, while the blue (light blue) lines represent the 5-day forecasts of GraphCast. The white square and circle denote the locations of maximum precipitation in the ERA5 reanalysis and GraphCast, respectively. In panel (c), the time axis has been shifted such that day 5 aligns with the day of maximum precipitation in ERA5 for each of the five events. The maximum precipitation for all the dynamically similar events in the training data (1979-2017) is marked with the grey dashed line.  
}
\label{fig: Fail-global-case}
\end{figure}
\FloatBarrier

To evaluate this possibility of extrapolating beyond the local events in the training data, we analyze GraphCast's performance on dynamically similar events from across the Northern Hemisphere (20\degree - 50\degree) with 12-hour accumulated precipitation higher than Dubai's. The highest precipitation from a extratropical cyclone in the ERA5-based \textit{training set} was 126~mm in 12 hours. We select the top five 12-hour accumulated precipitation events from the \textit{testing set} (2020-2023), which have maximum precipitation around 90 to 100~mm in 12 hours. Thus, none are global gray swans, but they are in-distribution at the very end of the tail (see Figure~\ref{fig: Dubai-PDF}b). One such event occurred in Florida in December 2023 (Figure~\ref{fig: Fail-global-case}a). This event, associated with a cyclone accompanied by frontal winds, is dynamically similar to the Dubai event. However, GraphCast's 5-day forecast significantly underestimates the rainfall, with peak values less than 60 mm (Figure~\ref{fig: Fail-global-case}a).

This underestimation of accumulated precipitation extremes is consistent across all five examined extreme events (Figure~\ref{fig: Fail-global-case}c and Figure~S4). While GraphCast has already seen even stronger dynamically similar events in the training set (with around 126~mm/12-hour precipitation rate), its 5-day forecasts have maxima between 30 and 60~mm/12-hour rates, much lower than the $\sim 90$~mm/12-hour rates of the actual events. 

\begin{figure}
\noindent\includegraphics[width=\textwidth]{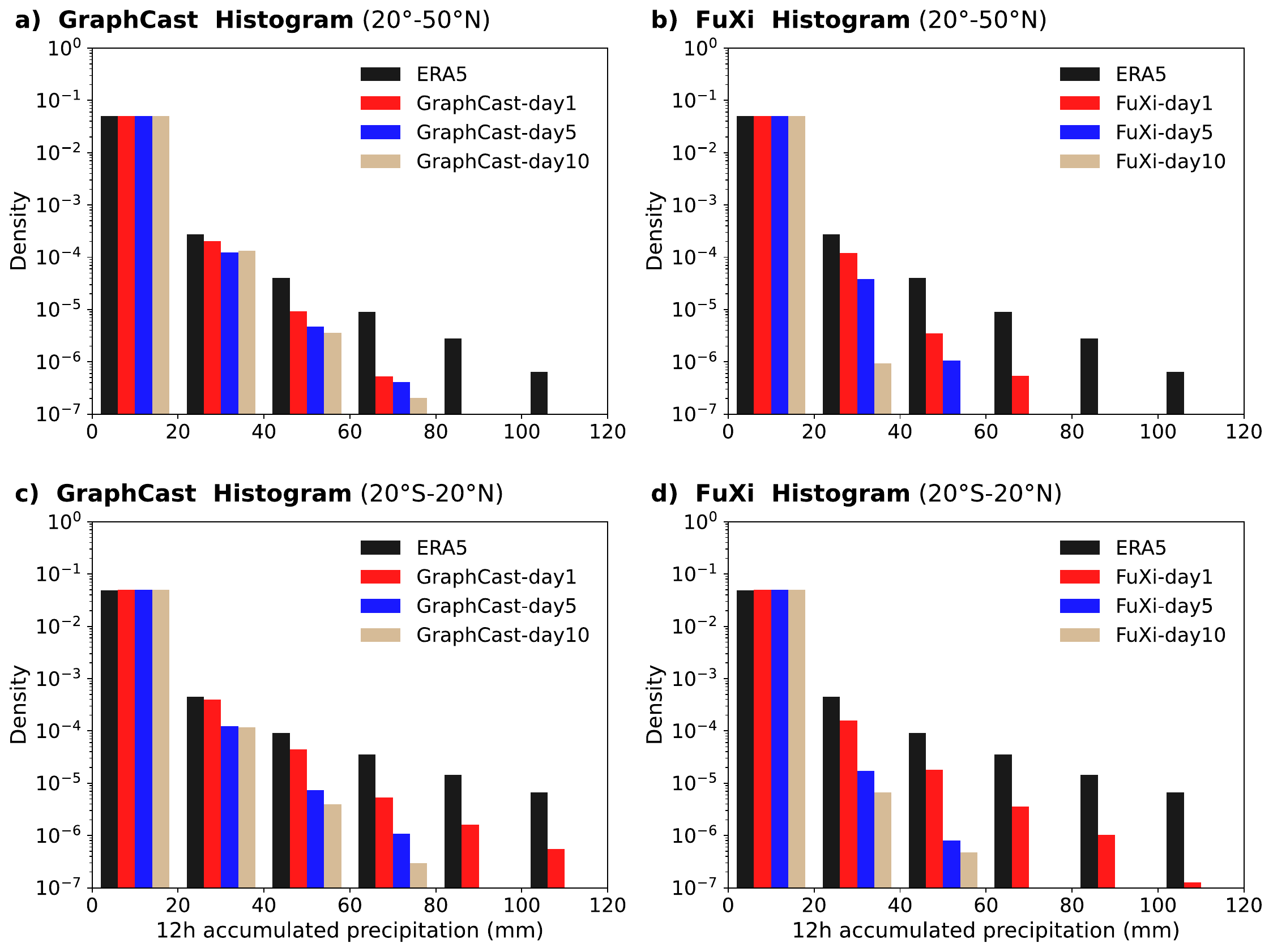}
\caption{\textbf{Poor representation of 12-hour accumulated precipitation distribution's tails in the 1-day, 5-day, and 10-day forecasts of GraphCast and FuXi.} Histograms represent aggregated forecasts initialized daily throughout April 2024. 
\textbf{(a)}-\textbf{(b)}:  GraphCast and Fuxi in the Northern Hemipshere latitude band 20\degree N--50\degree N;
\textbf{(c)}-\textbf{(d)}:  GraphCast and Fuxi in the tropical latitude band 20\degree S--20\degree N. Note that in these histograms include all events (not just dynamically similar events).}
\label{precip-histogram}
\end{figure}

The inability of GraphCast to reproduce the tail of the extratropical precipitation distribution can be clearly seen in the histograms of Figure~\ref{precip-histogram}. This shortcoming is even more pronounced in the tropics, where the precipitation amplitudes are higher. The same figure shows that the difficulty with reproducing the tail of the extratropical and tropical precipitation distribution is much more severe for FuXi. The problem worsens for both models at longer lead times (Figure~\ref{precip-histogram}); see Section~\ref{sec:fail} for further analysis and discussion.

So far, the results show that GraphCast and FuXi severely underestimate the strongest global extreme precipitation events that, while while rare, still fall within the distribution of the training data. This suggests that extrapolation (from local or global weaker events) is not a viable mechanism as these models even have challenges with infrequent yet in-distribution events. This might appear as strong evidence for a data imbalance problem. However, as briefly discussed next, at least some of this underestimation might be due to smoothing from spectral bias.


\section{Spectral bias as a limiting factor for the skill of AI models for extreme precipitation forecasting} \label{sec:fail}

\begin{figure}[ht]
\centering
\includegraphics[width=1.0\linewidth]{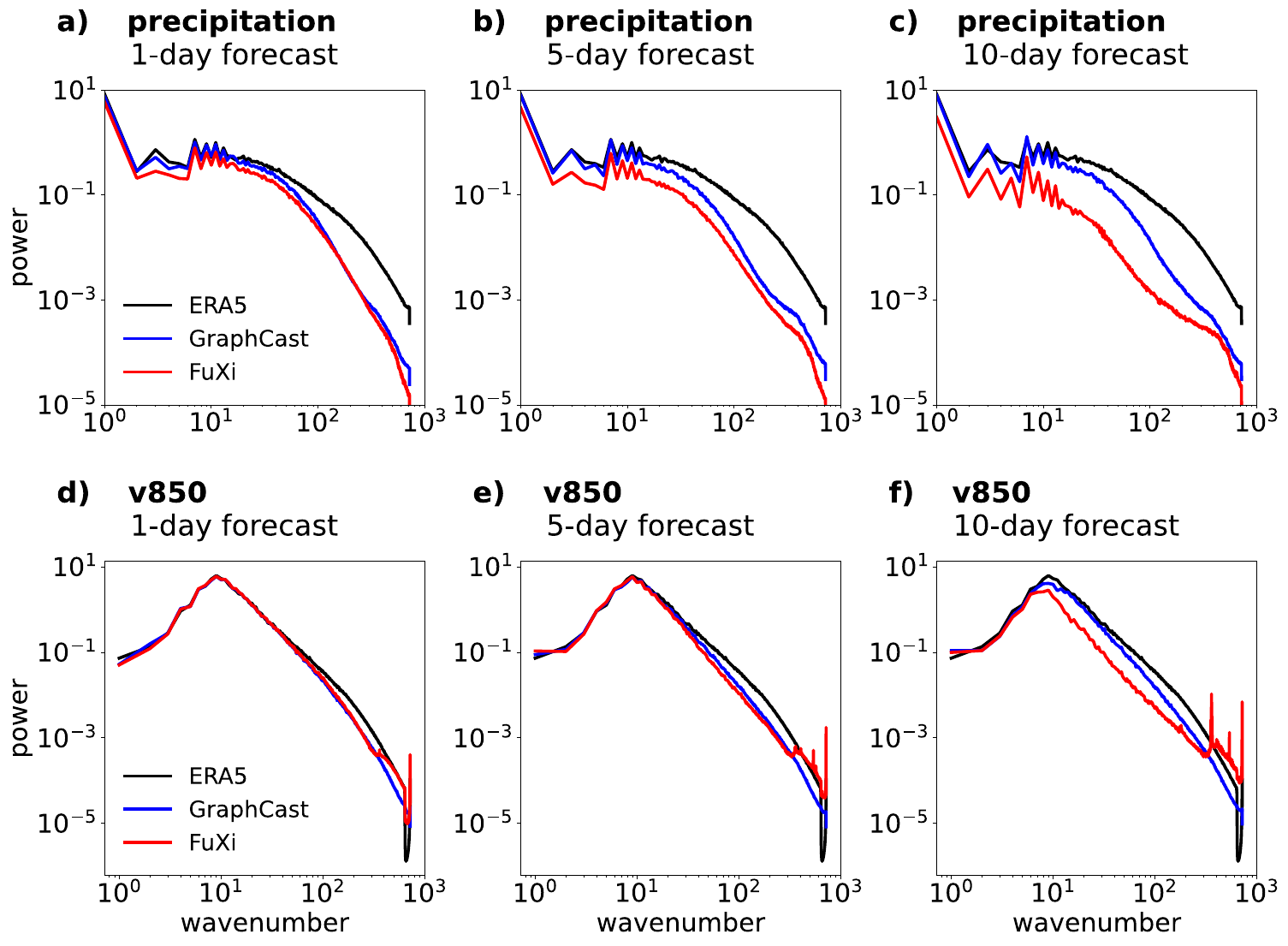}
\caption{\textbf{Poor representation of small scales in GraphCast and FuXi forecasts.} \textbf{(a)-(c)} Spectral power of 12-hour accumulated precipitation, and \textbf{(d)-(f)} meridional wind at 850 hPa (u850) as a function of wavenumber. The ground truth (ERA5 reanalysis) and forecasts from GraphCast and FuXi for different forecast lead times are compared. Spectral analysis using spherical harmonics is conducted for daily predictions of global fields in April 2024; hence, the results shown here are the average of 30 spectra.}
\label{fig: SB}
\end{figure}

To understand the difference in the performance of GraphCast and FuXi, we next examine the spectral bias of each model. 
As discussed in the Introduction, spectral bias \cite{rahaman2019spectral,xu2019frequency} is a common challenge for AI models, including the recent weather models \cite{chattopadhyay2023long, bonavita2024some, lai2024machine,subich2025fixing}: they preferentially learn large-scale features while struggling with the small-scale ones. Before discussing the root cause of this problem and its implications, we first show the representation of small scales (high wavenumbers) in the forecasts of GraphCast and FuXi, especially for precipitation. Figure~\ref{fig: SB}a-c presents the spectral power of 12-hour accumulated precipitation at 1, 5, and 10 days forecast lead times. At 1-day lead time, spectra of forecasts from both GraphCast and Fuxi show reasonable alignment with ERA5's at lower wavenumbers, which correspond to large-scale features, though with slightly lower spectral power (worse for FuXi). However, at higher wavenumbers (smaller scales), both models exhibit a noticeable underestimation of spectral power. This underestimation  indicates that even for short forecasts, both models struggle to capture finer-scale precipitation details.

As longer lead times of 5 and 10 days (Figure~\ref{fig: SB}b-c), the underestimation of spectral power in GraphCast's forecasts only slightly worsens. However, FuXi's spectra further deteriorate as the lead time increases, resulting in a significant underestimation of the power at small scales and even large scales on day 10 (panel c). The loss of spectral power across the scales can lead to the poor presentation of extreme events. This analysis explains FuXi's diminishing ability to capture the tail of the precipitation distribution as the lead time increases (Figure~\ref{precip-histogram}). To see the implication for Dubai's rainfall event, we have shown the time series for the 1-day forecast with FuXi in Figure~\ref{fig: Dubai-prediction}a.  Unlike the 5-day forecast, which only reached around 25~mm, the 1-day forecast is close GraphCast's 5-day forecast, in terms of onset and amplitude, peaking at 44~mm, above the 33~mm limit in the local training data. This suggests that the significantly underestimated peak in FuXI's 5-day forecast is largely attributable to the smoothing caused by spectral bias rather than due to the event's rarity (data imbalance). Note that spectral bias is also evident in the spectra of other variables, such as meridional wind at 850~hPa (Figure~\ref{fig: SB}d-f), but it is most pronounced for precipitation, which is dominated by small-scale features. This exacerbates the challenge of forecasting the inherently multi-scale processes where small-scale convection plays an important role.

The above analyses show that spectral bias is a major limiting factor in forecasting extreme precipitation events, especially at longer lead times. Before further discussions on why the magnitude of this bias can vary substantially between AI models, it is essential to explain its root causes.



Spectral bias, also known as the frequency principle, is a well-known problem in deep learning theory rooted in the optimization process during training~\cite{rahaman2019spectral, xu2019frequency}. Spectral bias is ubiquitous in applications of neural networks to multi-scale patterns, ranging from reconstruction of static images, e.g., of animals \cite{tancik2020fourier, mojgani2021closed}, to prediction of climate data \cite{chattopadhyay2023long,lai2024machine,mojgani2024interpretable,ng2024spectrum}. Chattopadhyay et al.~\cite{chattopadhyay2023long} discussed the theoretical and practical aspects of spectral bias in climate and weather applications in detail. 
Although spectral bias is usually seen as only the loss function problem (especially of mean-squared-error (MSE)-type losses), in fact, theoretical work shows that spectral bias strongly depends on the activation function as well \cite{xu2019frequency}.


Specific to weather and climate prediction, two additional factors contribute to spectral bias. First, in a chaotic atmosphere, small scales are less predictable than large scales. This issue, in particular, affects AI models that are trained with a few-day roll-out rather than with one time step, which includes both GraphCast (up to 3 days) and FuXi (up to 15 days). If the roll-out horizon exceeds the predictability limit, the model might attempt to suppress small-scale variability altogether, further reinforcing spectral bias. 

Second, when small-scale predictions do not precisely align with the observed spatial location, they incur a disproportionately high penalty—a phenomenon known as the ``double-penalty'' effect~\cite{ebert2013progress,subich2025fixing}. This discourages models from attempting to predict small-scale variations, leading to smoothing.

The more severe spectral bias in FuXi is likely stems from the model’s neural network architecture and optimization strategy, and in particular, the special adjustments to reduce MSE-type loss over longer lead times, which, as discussed above, can suppress the small scales. Overall, spectral bias can be a major hindrance to accurately forecasting the peak amplitude of extreme precipitation events. A number of general remedies for spectral bias in weather and climate models have emerged in recent years, e.g., via regularizing loss functions to better learn small scales, employing multi-stage training, and using diffusion models \cite{chattopadhyay2023long, mojgani2024interpretable, ng2024spectrum, kochkov2024neural, subich2025fixing,price2025probabilistic}. 

\FloatBarrier

\section{Summary and discussion} \label{sec:summary}

The key findings of the paper are
\begin{enumerate}
    \item GraphCast was able to forecast the location, timing, and the amplitude of the unprecedented 2024 Dubai rainfall and the associated circulation pattern remarkably well, even at an 8-day lead time (Figures~\ref{fig: Dubai-prediction} and S1). FuXi demonstrated similar skill, although it tended to significantly underestimate the rainfall amplitude except at much shorter lead times. 
    \item GraphCast's success  is particularly notable given that the event’s rainfall intensity and aspects of its dynamics, such as meridional moisture transport, were substantially out of the training distribution for that region (Figures~\ref{fig: Dubai-PDF} and S3). 
    \item GraphCast's success stems from translocation, the ability of the model to learn from dynamically similar events in other regions during training. This is enabled by the neural network's nearly global ERF (Figure~\ref{fig: ERF}). 
    \item The limited ability of GraphCast and FuXi to extrapolate from weaker events, data imbalance, and spectral bias are major limiting factors for the AI models in dealing with globally rare or unprecedented weather extremes~(Figures~\ref{fig: Fail-global-case}, \ref{precip-histogram}, and \ref{fig: SB}).
\end{enumerate}
Below, we further discuss (3) and (4) and their practical and theoretical implications for AI weather models and the emerging AI emulators of long-term atmospheric and oceanic variability built on them \cite{watt2024ace2, dheeshjith2024samudra, chapman2025camulator}. 



\textbf{Success via translocation and its implications}: We show that GraphCast has a nearly global ERF, which means that, during both training and testing, even for one 6-hourly time-step forecast over a specific region like Dubai, the model draws on information from the entire globe (Figure~\ref{fig: ERF}). At the same time, we demonstrate that while the Dubai event's rainfall and some dynamical aspects were significantly out-of-distribution for that region, they were fairly in-distribution among events with similar dynamics \textit{globally} (Figure~\ref{fig: Dubai-PDF}). Taken together and combining our results with the insight from the controlled experiments of Sun~et al.~\cite{sun2024can}, we infer that the success of GraphCast in forecasting this regionally unprecedented event results from translocation. Note that while we did not compute the ERF of FuXi due to computational constraints, the ERF of deep, large neural networks such as the one used in FuXi is expected to be nearly global~\cite{pahlavan2024importance}. 

The ability of these AI models to learn from events in other regions is encouraging from a practical perspective. It suggests that the models may be capable of short-term forecasting and long-term emulation of regionally unprecedented events that arise from natural variability or shifting weather patterns due to climate change, provided that similar events have occurred elsewhere in the training data. Moreover, this transferability may offer significant value for regions with limited historical observations, such as Africa’s sparse precipitation records~\cite{bassine2025role}. Nevertheless, more work is needed to systematically evaluate the performance of AI weather and climate models on these fronts. Tools like ERFs and controlled experiments, such as those in~\cite{sun2024can, hakim2024dynamical,van2025reanalysis}, offer promising pathways for advancing this understanding.

From a theoretical perspective, the ability to translocate suggests that AI models learn generalizable physical relationships and statistical structures, rather than overfitting to location-specific data. This adds to the growing evidence, e.g., from a few recent studies with perturbation experiments and adjoints~\cite{hakim2024dynamical, bano2025ai,van2025reanalysis}, that AI weather models learn physically meaningful spatio-temporal relationships; however, as further discussed below, this learning does not seem to suffice for extrapolation.

An interesting question is how translocation can be further improved and exploited in the design of the next generation of AI weather and climate models. Recently, ERFs have been discussed in a few studies on weather/climate modeling. Most notably, Karlbauer~et al.~\cite{karlbauer2024advancing} estimated the physically meaningful size of receptive fields in AI models of atmosphere based on the speed of sound ($\sim 300$~m/s) and maximum speed of the jet stream winds ($\sim 100$~m/s), which for a 6-hourly time step, yield around 6500~km and 2150~km in each direction, respectively. Along the same lines of connecting the ERF size with atmospheric dynamics, Pahlavan~et al.~\cite{pahlavan2024importance} examined various emulators of a 1D model of quasi-biennial oscillation (QBO), concluding that extensively non-local ERFs are needed to capture nearly instantaneous long-range correlations that can arise from filtered fast processes such as gravity waves. The results of the current study, however, suggest a potentially critical benefit in having large ERFs, beyond what would be physically meaningful spatial scales, to enable translocation. Nonetheless, large ERFs can also increase the likelihood of learning spurious and unphysical relationships, warranting further investigation into the best choices for ERFs. 

\textbf{Limiting factors and the need for innovations}: The three aforementioned limiting factors (lack of extrapolation, spectral bias, and data imbalance) have different drivers and likely require different innovative solutions. Addressing spectral bias has received considerable attention in recent years~\cite{chattopadhyay2023long, mojgani2024interpretable, ng2024spectrum, kochkov2024neural, subich2025fixing, chakraborty2025binned, khodakarami2025mitigating}. In particular, the use of diffusion models, e.g., in GenCast~\cite{price2025probabilistic}, appears to significantly reduce the spectral bias. Future work should further investigate to what degree these approaches improve the representation of the dynamics (rather than just statistics) of the small-scale processes.

Data imbalance for in-distribution events can be potentially mitigated by weighting the loss function~\cite{miloshevich2023probabilistic,rudy2023output,sun2024data,yang2024overcoming}. For globally out-of-distribution events, a promising approach is oversampling, i.e., generating stronger extreme events using physics-based models guided by techniques such as rare event sampling (RES) algorithms~\cite{ragone2018computation,ragone2021rare,webber2019practical,finkel2024bringing} and ensemble boosting~\cite{fischer2023storylines} for the re-training of the AI model. Note that AI models themselves can accelerate RES algorithms~\cite{Lancelin2025}, thus, building a feedback loop between RES, AI model, and the physics-based model. 

Finally, the question of extrapolation from weaker events requires extensive and careful investigations. In particular, one has to disentangle extrapolation and translocation, as for example, in the case of the Dubai event, what might appear as extrapolation is in fact translocation. Thus, studies of extrapolation should focus on events that are at the tail of, or out of, the global distribution of the training set. Note that the analyses of this paper or \cite{sun2024can} do not exclude the possibility that the models might be able to do some extrapolation. Results of \cite{sun2024can} shows lack of significant extrapolation (from Category 2 to 5 TCs), while any potential extrapolation in the current study could have been dwarfed by spectral bias. Regardless, innovations in architecture and training algorithms might be needed to further improve the ability of these models in learning physics, to the degree that extrapolation might become possible.

In conclusion, rapidly advancing AI weather models already demonstrate skill in forecasting unprecedented events such as the 2024 Dubai rainfall. Advancing our understanding of the mechanisms underlying their successes and limitations, and disentangling the sources of their errors, can help unlock the full potential of AI-driven weather and climate models for both accurate prediction and scientific discovery.

\section*{Methods}

\subsection*{AI weather models and ERA5 data}

At the time of conducting the analyses in this study, only two already trained AI weather models, GraphCast and FuXi, directly forecasted precipitation. The code used to produce forecasts from the trained GraphCast is available at \url{https://github.com/google-deepmind/graphcast}. GraphCast has multiple versions with different numbers of vertical levels; in this study, we use a 13-level configuration to generate 10-day forecasts at 0.25° spatial resolution from 6-hourly autoregressive roll-outs. The model was trained on ERA5 reanalysis data from 1979 to 2017, and was additionally fine-tuned on a smaller sample of IFS HRES data from 2016 to 2021. The same model can also be accessed publicly at \url{https://github.com/ecmwf-lab/ai-models-graphcast}.

The model weights used to produce FuXi forecasts are available at \url{https://github.com/tpys/FuXi}. FuXi was also trained with ERA5 reanalysis data (1979-2015 for training and 2016-2017 for validation) to generate 6-hourly forecasts up to 15 days ahead at a spatial resolution of 0.25°. A key difference of FuXi is that it employs a cascade of models trained for three sequential forecast periods of 0–5 days, 5–10 days, and 10–15 days, respectively, to optimize its performance for short- and medium-range weather forecasting.

We also used Pangu-Weather and FourCastNet-V2 for circulation forecasts (see Figure S2). Both models are publicly available at \url{https://github.com/ecmwf-lab/ai-models-panguweather} and \url{https://github.com/ecmwf-lab/ai-models-fourcastnetv2}.

The ERA5 reanalysis dataset~\cite{hersbach2020era5} is used as ground-truth to test the AI weather models in this study. The dataset is publicly available through the Copernicus Climate Data Store (CDS) at \url{https://cds.climate.copernicus.eu/}. The ERA5, in general, may not capture the highest observed precipitation totals~\cite{lavers2022evaluation}. However, since AI models are trained with ERA5 data, we are testing the models in the world of ERA5 to better understand their potential and limitations.

\subsection*{Identifying dynamically similar events using Lagrangian tracking}

The Lagrangian detection and tracking algorithm described in Crawford et al. \cite{crawford2021sensitivity} is used  to track extratropical cyclones in the ERA5 dataset. This algorithm identifies and tracks local minima in sea level pressure (slp), where ``local'' is defined within the 24 nearest neighbors (i.e., within a $5 \times 5$ box). The slp fields are analyzed at a spatial resolution of 50~km and a temporal resolution of 3~hours. The tracking code is publicly available at \url{https://github.com/alexcrawford0927/cyclonetracking}.

To ensure that tropical storms are excluded from the analysis, TCs in the International Best Track Archive for Climate Stewardship (IBTrACS) dataset are compared with the cyclone centers identified by the tracking algorithm. If a TC from IBTrACS is located within 500~km of a tracked event, that event is considered to be influenced by the TC and is excluded from the ``Tropical cyclones removed'' historam of Figure~\ref{fig: Dubai-PDF}b. 

After removing events influenced by TC, dynamically similar events are then defined as those occurring during the boreal winter (November to April) and associated with a maximum precipitation rate within 1500 km of the cyclone center exceeding 60 mm/12-hour. The histogram of precipitation rates for ``Dynamically similar events'' is shown in Figure~\ref{fig: Dubai-PDF}b. Note that a single selected event can contribute multiple data points to the histogram, corresponding to precipitation rates ranging from 1 mm/12-hour to 100 mm/12-hour.

\subsection*{Calculation of effective receptive field (ERF)}

To quantify the extent to which the neural network of an AI weather model uses spatially non-local information for prediction of a variable at a given location, we use the effective receptive field (ERF), which can be numerically calculated for any  architecture~\cite{luo2016understanding,pahlavan2024importance}. The ERF measures how much a specific output is influenced by small changes in the input, identifying the regions of the input that are more influential. Calculating the ERF involves backpropagating a signal from the output layer to the input layer, similar to techniques such as saliency maps~\cite{simonyan2013deep} and layer-wise relevance propagation (LRP)~\cite{montavon2018methods}, which have been recognized as effective visualization tools in climate applications~\cite{toms2020physically,ebert2020evaluation,chen2024machine}.

To calculate the ERF of GraphCast for predicting precipitation over Dubai for a single time step (6 hours), we initialize the model at 12:00 UTC, April 15, 2024. Within the Dubai region (defined as 23N°–28N° latitude and 50E°–60E° longitude), we identify the grid cell with the strongest precipitation for the next forecast at time $t$, \(P_{\text{Dubai}}=P(y=\text{Dubai},t)\). $y$ represents location (latitude, longitude, and altitude) . Note that $P(y,t)$ is one of the variables in the state vector $X(y,t)$, the input and output of the GraphCast's neural network $\mathcal{N}$:
\begin{equation}
X(y,t+1)=\mathcal{N}(X(y,t-1),X(y,t),\theta),
\end{equation}
\noindent where $\theta$ represent the weights of the neural network. We compute the partial derivative \(\partial{P_{\text{Dubai}}}/\partial{X}\). This partial derivative is calculated by backpropagating a gradient from the output to the input of $\mathcal{N}$, similar to the standard process of propagating the error gradient relative to a loss function~\cite{luo2016understanding}. Following previous studies~\cite{luo2016understanding}, the gradient is set to zero for all outputs except the specific point of interest, \(P_{\text{Dubai}}\), allowing the ERF to reveal the sensitivity of this output to variations in the input.

\bibliography{ref}

\section*{Acknowledgements}
We thank Michael Kremer of UChicago's Economics Department for suggesting this test case. This work was supported by NSF grant AGS-2046309 and the University of Chicago's Data Science Institute 
(DSI) through the AI for Climate Initiative. Computational resources were provided by NSF ACCESS (allocation ATM170020) and NCAR's CISL (allocation URIC0009).

\section*{Author contributions statement}

Y.S. and P.H. conceived the research. Y.S. and H.P. conducted analyses. All authors interpreted the results. Y.S., H.P., P.H., and T.S. wrote the paper. All authors edited the manuscript. 

\section*{Competing interests}
The authors declare no competing interests.

\end{doublespace}

\newpage


\renewcommand{\thefigure}{S\arabic{figure}}
\renewcommand{\thetable}{S\arabic{table}}
\renewcommand{\theequation}{S\arabic{equation}}
\setcounter{figure}{0}
\setcounter{table}{0}
\setcounter{equation}{0}


\begin{doublespace}
\nolinenumbers  

\vspace{50mm}
\section*{Supporting Information for \\ ``Predicting Beyond Training Data via Extrapolation versus Translocation: AI Weather Models and Dubai's Unprecedented 2024 Rainfall''}


\vspace{15mm}

\author{
Y. Qiang Sun\textsuperscript{1,*},
Pedram Hassanzadeh\textsuperscript{1,*},
Tiffany Shaw\textsuperscript{1},
Hamid A. Pahlavan\textsuperscript{2}
}

\noindent\textsuperscript{1}University of Chicago, Department of the Geophysical Sciences, Chicago, IL 60637, USA\\
\textsuperscript{2}NorthWest Research Associates, Boulder, CO 80301, USA\\
\textsuperscript{*}Corresponding authors: qiangsun@uchicago.edu and pedramh@uchicago.edu

\end{doublespace}


\vspace{10mm}
\nolinenumbers
\subsubsection*{This PDF file includes:}
Figures S1 to S4\\

\newpage

\begin{figure}
\noindent\includegraphics[width=\textwidth]{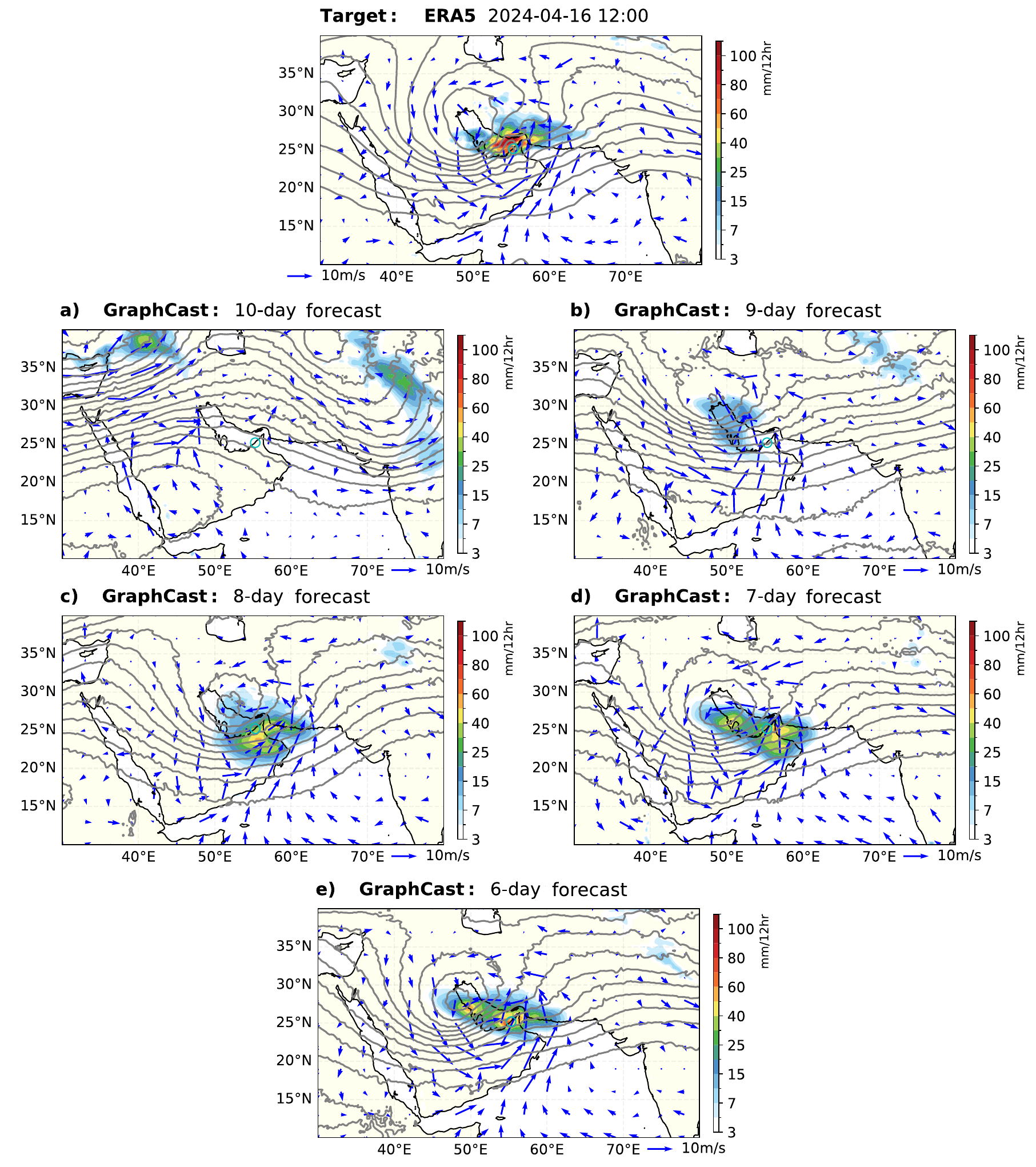}
\caption{\textbf{GraphCast forecasts of the Dubai event at various lead times.} 
\textbf{a)} 10 days; 
\textbf{b)} 9 days; 
\textbf{c)} 8 days lead time, note the correct location and the precipitation amplitudes that exceed the training maximum near Dubai already; 
\textbf{d)} 7 days; 
\textbf{e)} 6 days.}

\label{SI-8day}
\end{figure}

\begin{figure}
\noindent\includegraphics[width=\textwidth]{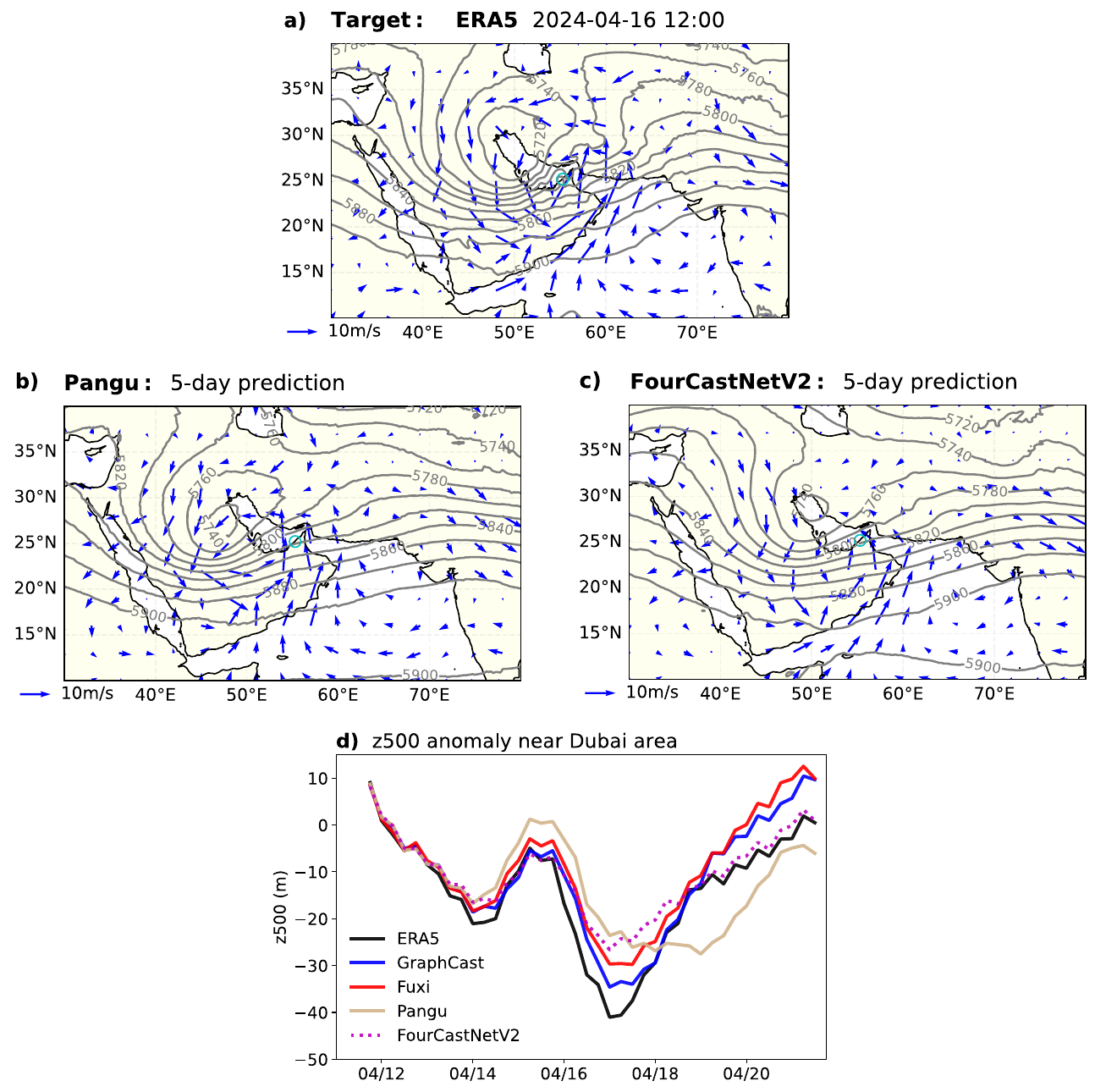}
\caption{\textbf{Forecasts by the Pangu-Weather and FourCastNetV2 models for the event.} 
\textbf{a)} Geopotential height at 500 hPa (z500) and meridional wind (v850) from ERA5 reanalysis; 
\textbf{b)} Forecast from the Pangu-Weather model, initialized on April 11th; 
\textbf{c)} Forecast from the FourCastNetV2 model, initialized on April 11th; 
\textbf{d)} Time evolution of the area-averaged geopotential height over the Dubai region as represented in ERA5 and in forecasts produced by the four AI models in this study.}

\end{figure}

\begin{figure}
\noindent\includegraphics[width=\textwidth]{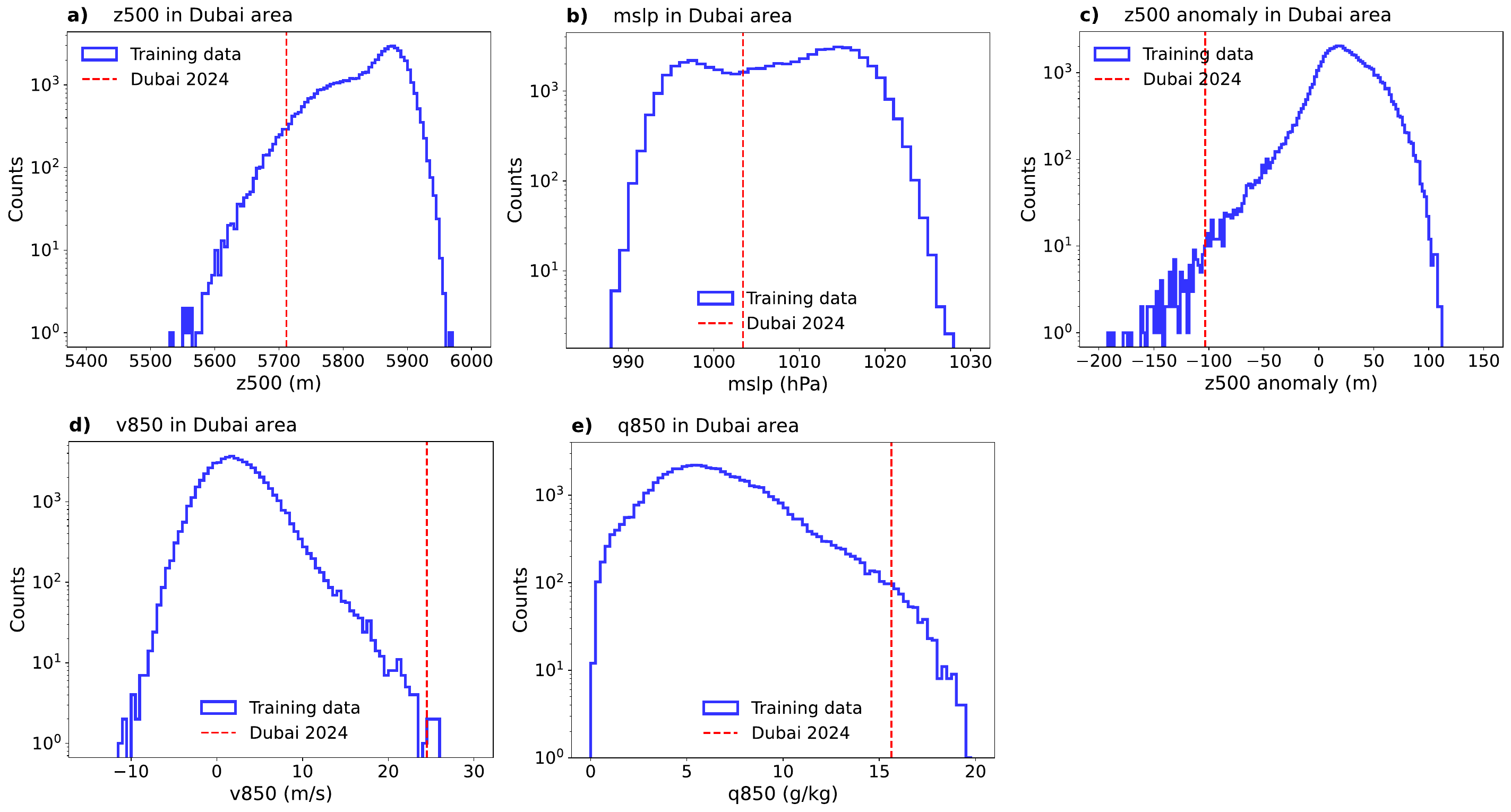}
\caption{\textbf{Rarity of the Dubai event in different variables.} \textbf{(a)} Geopotential height at 500 hPa (z500); \textbf{(b)} Sea level pressure (slp); \textbf{(c)} Geopotential height anomaly at 500 hPa (z500 anomaly); \textbf{(d)} Meridional wind at 850 hPa (v850); \textbf{(e)} Specific humidity at 850 hPa (q850). When examined individually, none of these variables is OOD. However, when considered together, some of the joint distributions, most notably that of v850-q850, are OOD, signifying, for example, unprecedented meridional moisture transport  (Figure 2c in the main text).}
\label{SI-OOD}
\end{figure}

\begin{figure}
\noindent\includegraphics[width=\textwidth]{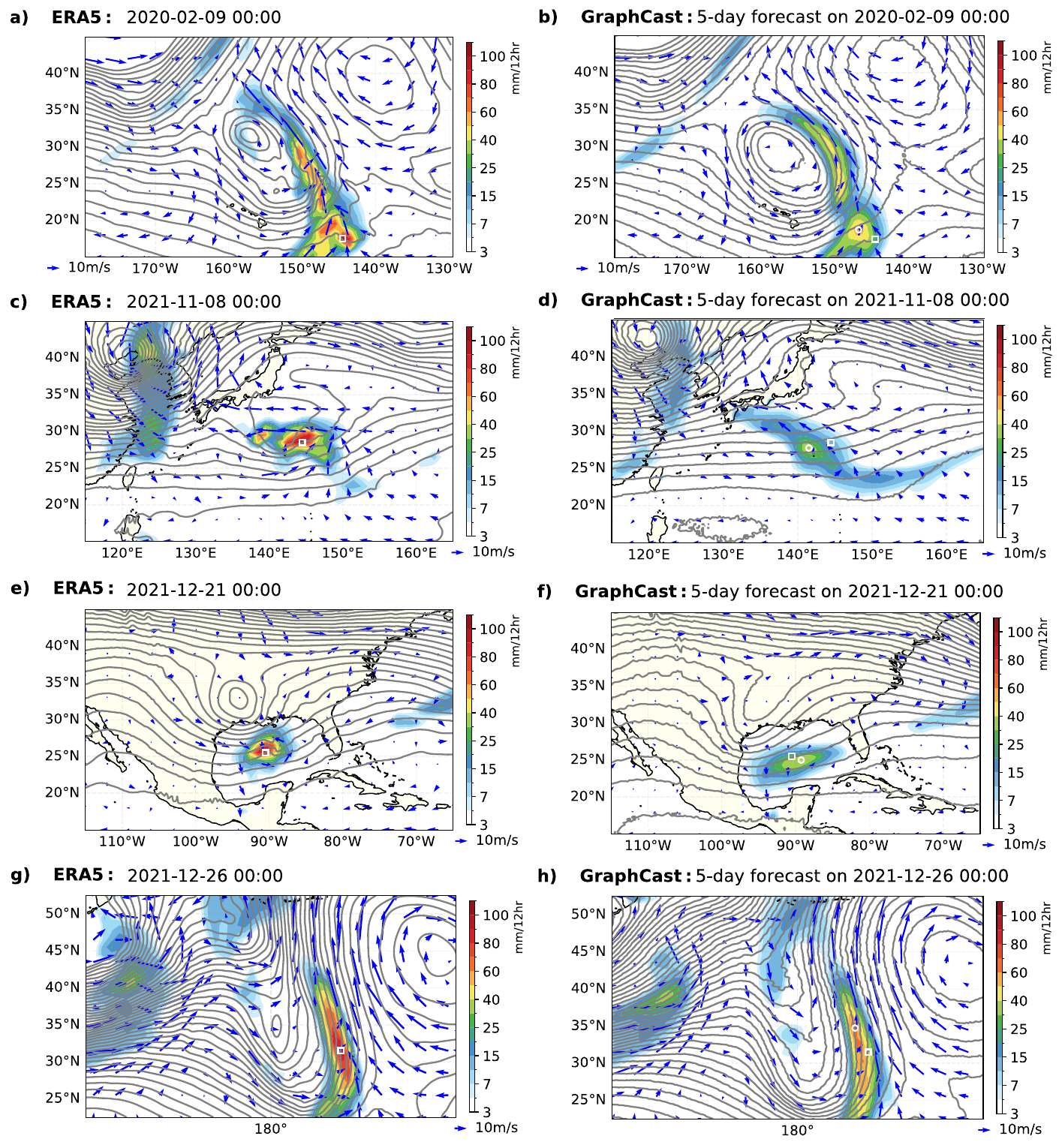}
\caption{\textbf{GraphCast forecasts of global extreme events.} Similar to Figure 4 in the main text, this figure presents four additional cases. The left column shows ERA5 reanalysis data for extreme rainfall events, while the right column displays GraphCast's 5-day forecasts for the same fields. In all cases, the predicted precipitation is significantly weaker than the observed values.}
\label{SI-global-extreme}
\end{figure}


\end{document}


%
%


\title{Supporting Information for ``Predicting Beyond Training Data via Extrapolation versus Translocation: AI Weather Models and Dubai's Unprecedented 2024 Rainfall''}
%
%

%
%



\author{
Y. Qiang Sun\textsuperscript{1,*},
Pedram Hassanzadeh\textsuperscript{1,*},
Tiffany Shaw\textsuperscript{1},
Hamid A. Pahlavan\textsuperscript{2}
}

\noindent\textsuperscript{1}University of Chicago, Department of the Geophysical Sciences, Chicago, IL 60637, USA\\
\textsuperscript{2}NorthWest Research Associates, Boulder, CO 80301, USA\\
\textsuperscript{*}Corresponding authors: qiangsun@uchicago.edu or pedramh@uchicago.edu




%
%

%

\begin{article}

%
%

\noindent\textbf{Contents of this file}
\begin{enumerate}

\item Figures S1 to S4

\end{enumerate}

\end{article}


%
%
%
%
%
%
%
%

\begin{figure}
\noindent\includegraphics[width=\textwidth]{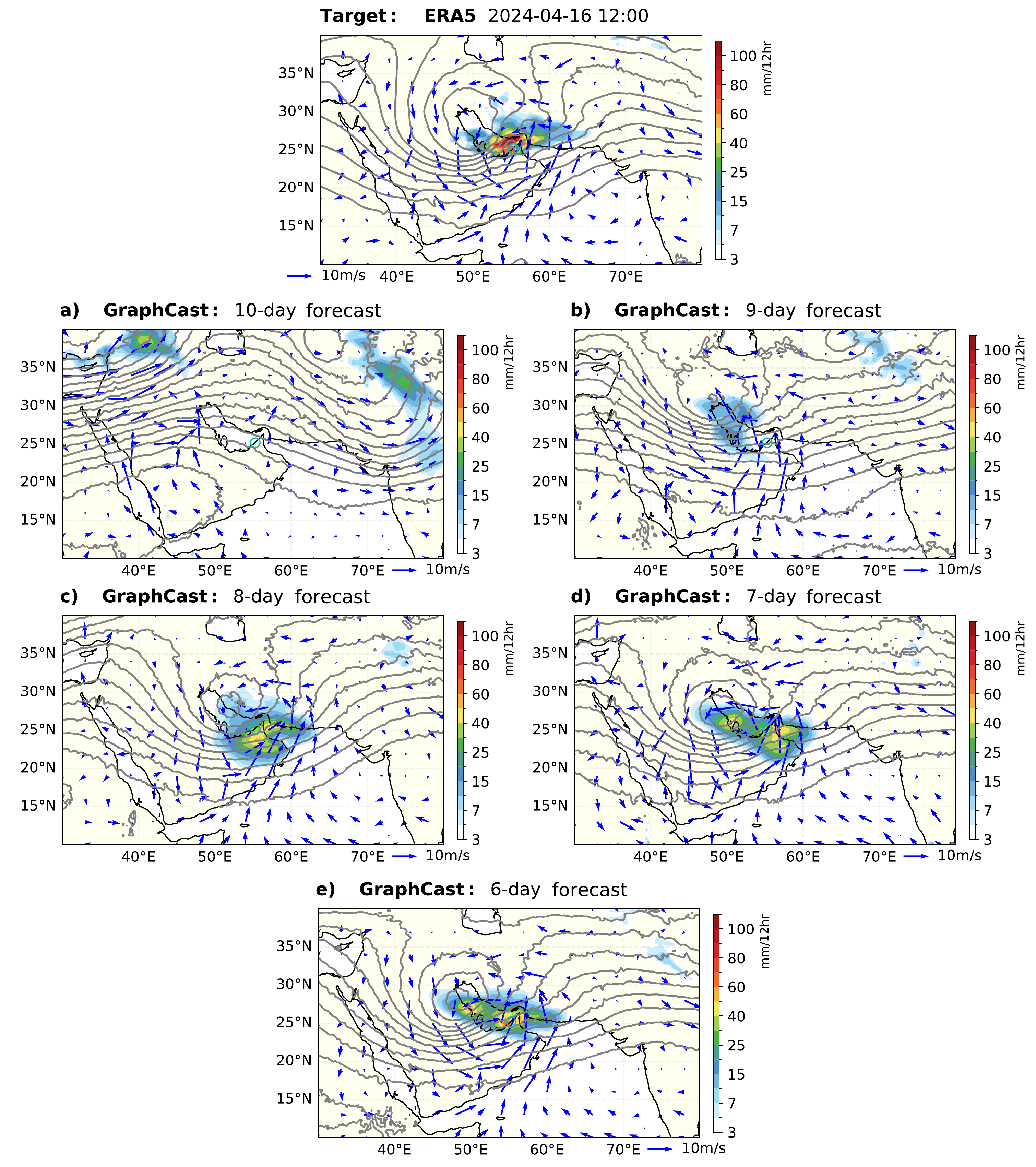}
\caption{\textbf{GraphCast forecasts of the Dubai event at various lead times.} 
\textbf{a)} 10 days; 
\textbf{b)} 9 days; 
\textbf{c)} 8 days lead time, note the correct location and the precipitation amplitudes that exceed the training maximum near Dubai already; 
\textbf{d)} 7 days; 
\textbf{e)} 6 days.}

\label{SI-8day}
\end{figure}

\begin{figure}
\noindent\includegraphics[width=\textwidth]{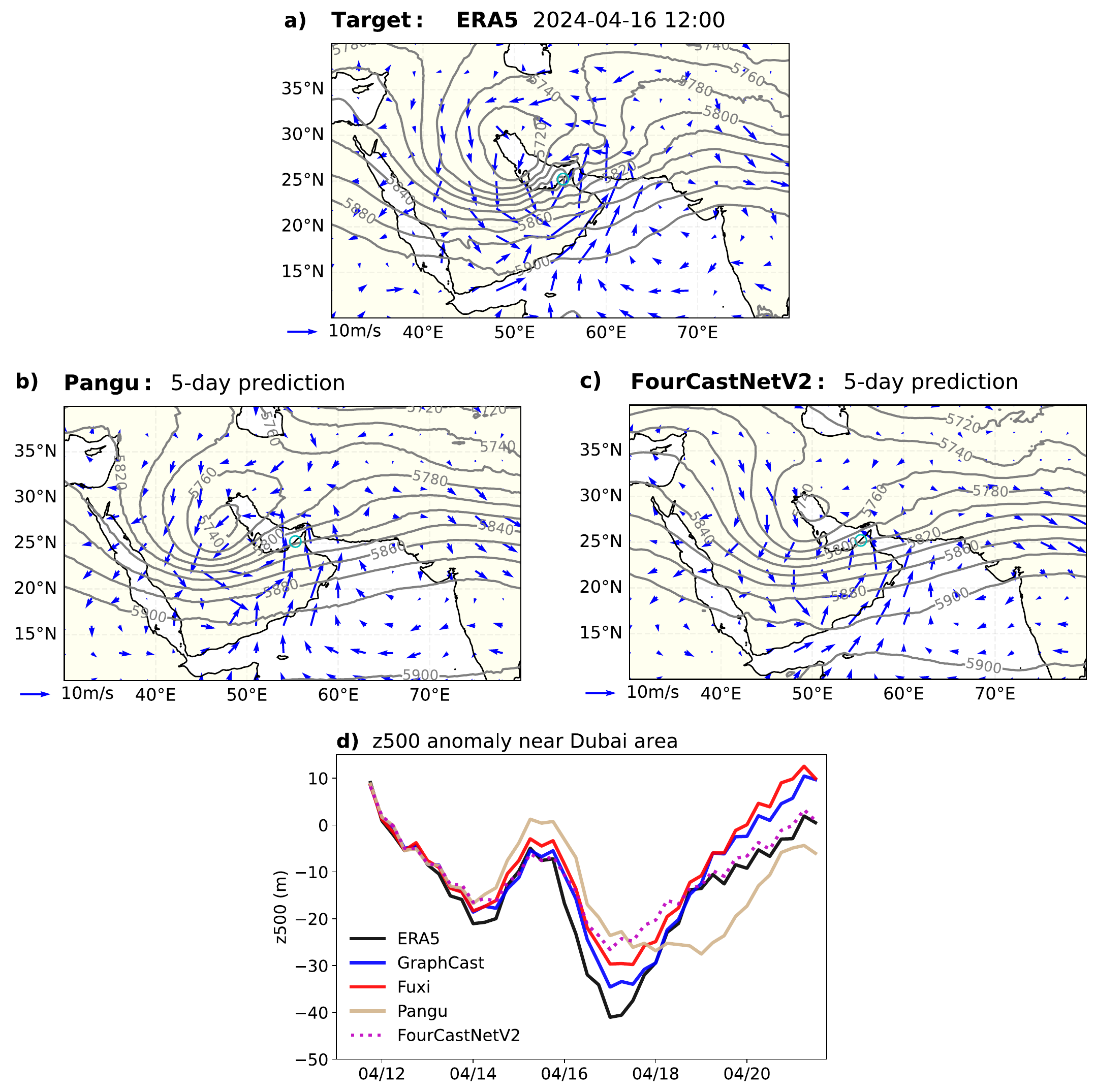}
\caption{\textbf{Forecasts by the Pangu-Weather and FourCastNetV2 models for the event.} 
\textbf{a)} Geopotential height at 500 hPa (z500) and meridional wind (v850) from ERA5 reanalysis; 
\textbf{b)} Forecast from the Pangu-Weather model, initialized on April 11th; 
\textbf{c)} Forecast from the FourCastNetV2 model, initialized on April 11th; 
\textbf{d)} Time evolution of the area-averaged geopotential height over the Dubai region as represented in ERA5 and in forecasts produced by the four AI models in this study.}

\end{figure}

\begin{figure}
\noindent\includegraphics[width=\textwidth]{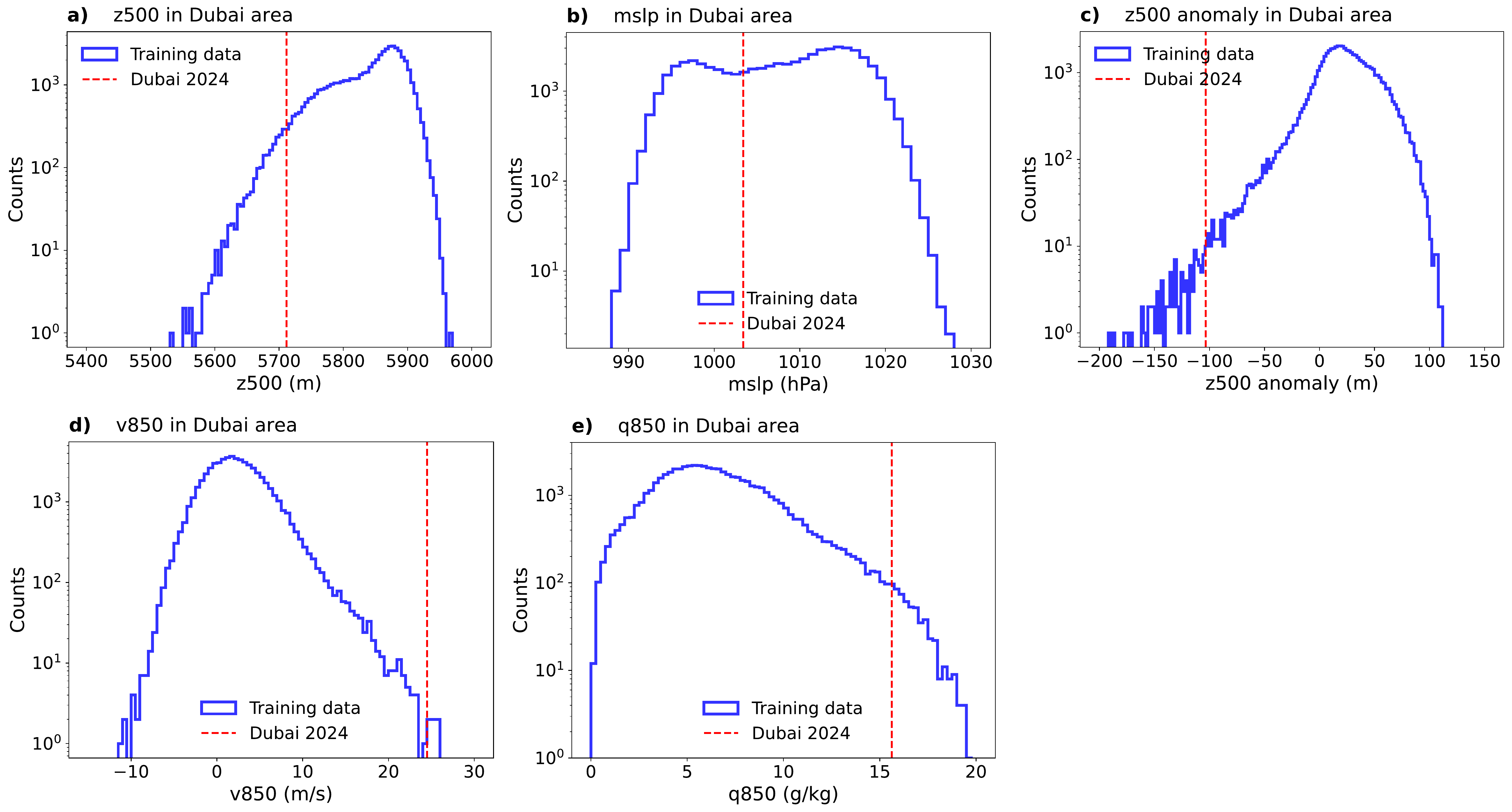}
\caption{\textbf{Rarity of the Dubai event in different variables.} \textbf{(a)} Geopotential height at 500 hPa (z500); \textbf{(b)} Sea level pressure (slp); \textbf{(c)} Geopotential height anomaly at 500 hPa (z500 anomaly); \textbf{(d)} Meridional wind at 850 hPa (v850); \textbf{(e)} Specific humidity at 850 hPa (q850). When examined individually, none of these variables is OOD. However, when considered together, some of the joint distributions, most notably that of v850-q850, are OOD, signifying, for example, unprecedented meridional moisture transport  (Figure 2c in the main text).}
\label{SI-OOD}
\end{figure}

\begin{figure}
\noindent\includegraphics[width=\textwidth]{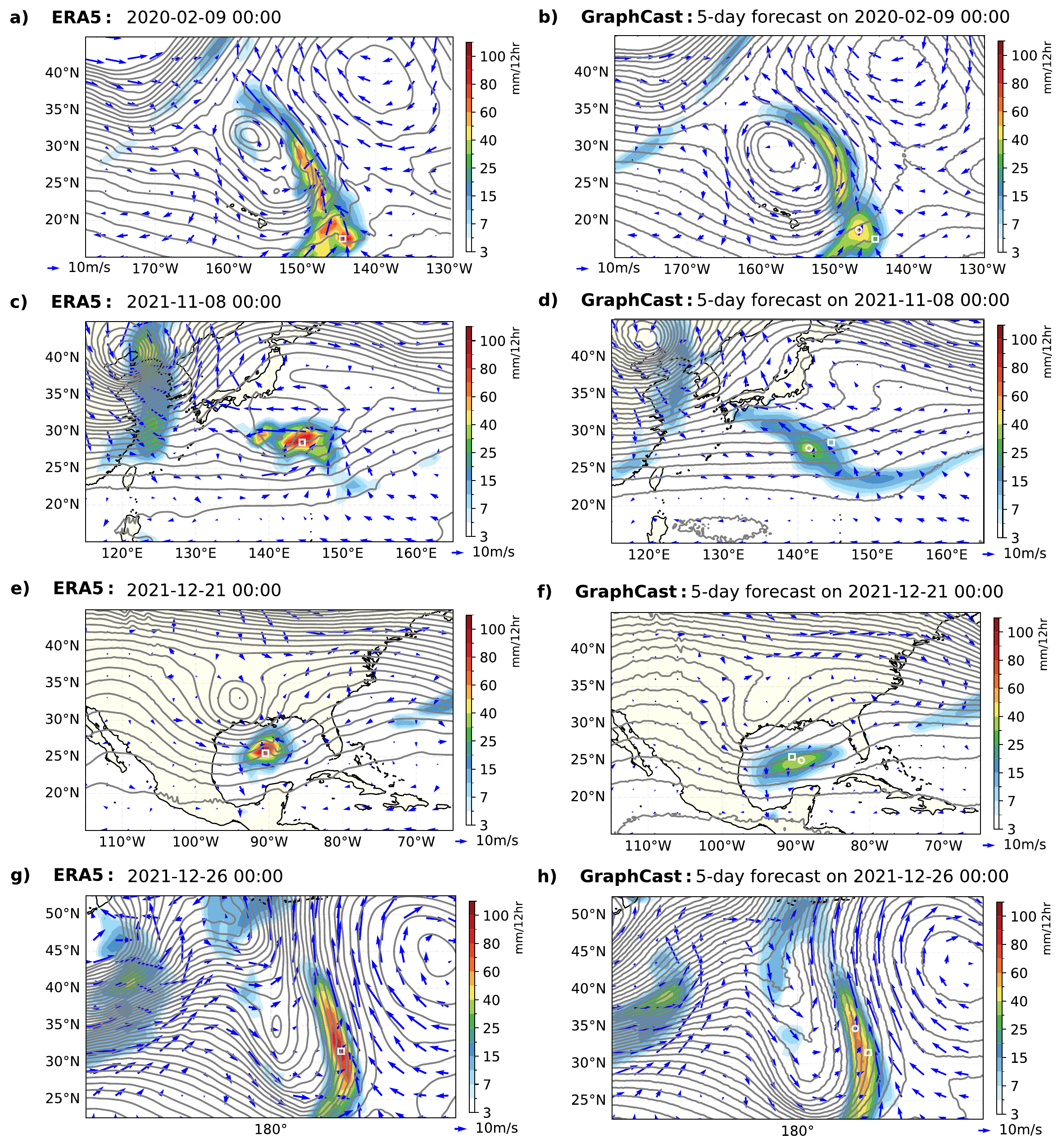}
\caption{\textbf{GraphCast forecasts of global extreme events.} Similar to Figure 4 in the main text, this figure presents four additional cases. The left column shows ERA5 reanalysis data for extreme rainfall events, while the right column displays GraphCast's 5-day forecasts for the same fields. In all cases, the predicted precipitation is significantly weaker than the observed values.}
\label{SI-global-extreme}
\end{figure}


%
%
%
%
%